\documentstyle[11pt]{article}
\newcommand{\R}{{\sf R\hspace*{-0.9ex}\rule{0.15ex}%
{1.5ex}\hspace*{0.9ex}}}
\def\bkR{{\rm I\kern-.17em R}}
\def\bkC{{\rm \kern.24em \vrule width.05em height1.4ex depth-.05ex \kern-.26em C}}

\begin{document}

\title{Exact master equation for a noncommutative Brownian particle}
\author{Nuno Costa Dias\footnote{{\it ncdias@meo.pt}} \\ Jo\~{a}o Nuno Prata\footnote{{\it joao.prata@ulusofona.pt}} \\ {\it Departamento de Matem\'atica} \\
{\it Universidade Lus\'ofona de Humanidades e Tecnologias} \\ {\it Av. Campo Grande, 376, 1749-024 Lisboa, Portugal}\\
{\it and}\\
{\it Grupo de F\'{\i}sica Matem\'atica}\\
{\it Universidade de Lisboa}\\
{\it Av. Prof. Gama Pinto 2}\\
{\it 1649-003, Lisboa, Portugal}}

\maketitle

\maketitle

\begin{abstract}
We derive the Hu-Paz-Zhang master equation for a Brownian particle linearly coupled to a bath of harmonic oscillators on the plane with spatial noncommutativity. The results obtained are exact to all orders in the noncommutative parameter. As a by-product we derive some miscellaneous results such as the equilibrium Wigner distribution for the reservoir of noncommutative oscillators, the weak coupling limit of the master equation and a set of sufficient conditions for strict purity decrease of the Brownian particle. Finally, we consider a high-temperature Ohmic model and obtain an estimate for the time scale of the transition from noncommutative to ordinary quantum mechanics. This scale is considerably smaller than the decoherence scale.
\end{abstract}


\newpage

\section{Introduction}

The question of space-time noncommutativity has a long-standing story. It was put forward by Snyder \cite{Snyder}, Heisenberg, Pauli \cite{Heisenberg} and Yang \cite{Yang} as a means to regularize quantum field theories. However, the development of renormalization techniques and certain undesirable features of noncommutative theories such as the breakdown of Lorentz invariance have hindered further research in this direction. More recently, several important developments in various approaches to the quantization of gravity have revived the interest in the concept of noncommutative space-time. See for instance \cite{Freidel} in the context of 3d gravity. In the realm of string theory, the discovery that the low energy effective theory of a D-brane in the background of a Neveu-Schwarz-Neveu-Schwarz B field lives on a space with spatial noncommutativity \cite{Douglas}, \cite{Hull}, \cite{Seiberg} has triggered an enormous amount of research in this field. From another perspective, a simple heuristic argument, based on Heisenberg's uncertainty principle, the equivalence principle and the Schwarzschild metric, shows that the Planck length seems to be a lower bound on the precision of a measurement of position \cite{Rosenbaum}. This reenforces the point of view that a new (noncommutative) geometry of space-time will emerge at a fundamental level. The simplest way to implement these ideas in quantum field theories is by adding to the phase-space noncommutativity of quantum mechanics a new space-time noncommutativity. Connes' noncommutative geometry \cite{Connes}, \cite{Martinetti} provides a suitable mathematical framework for the formulation of quantum field theories in noncommutative space-time. These theories have been intensively studied in the literature (for a review see \cite{Nekrasov}, \cite{Szabo}). In most of the approaches \cite{Martinetti}, \cite{Guisado} the space-time coordinates $x^{\mu}$ do not commute, either in a canonical way
\begin{equation}
\left[x^{\mu}, x^{\nu} \right] = i \theta^{\mu \nu},
\end{equation}
or in a Lie-algebraic way
\begin{equation}
\left[x^{\mu}, x^{\nu} \right] = i C_{\beta}^{\mu \nu} x^{\beta},
\end{equation}
where $\theta^{\mu \nu}$ and $C_{\beta}^{\mu \nu}$ are constants. In general, one assumes that $\theta^{0i}=0$ (or $C_{\beta}^{0i}=0$) (i.e. that time is an ordinary commutative parameter) to avoid problems with the lack of unitarity.\footnote{This claim seems debatable. See \cite{Doplicher} for an alternative.} Equations (1,2) can be regarded as a particular case of the more general deformation $\left[x^{\mu}, x^{\nu} \right] = i f^{\mu \nu} (x)$ \cite{Bertolami, Carroll}, where $f^{\mu \nu} (x)$ are real functions of $x$ such that $f^{\mu \nu} (x)= - f^{\nu \mu} (x)$\footnote{see \cite{Daszkiewicz} for a recent survey of the various types of deformations and some of there consequences at the classical level.}. This occurs when the laboratory frame has a space-time dependent motion. As proposed in \cite{Kamoshita} it is likely that $f^{\mu \nu}$ has a fixed value in the Cosmic Microwave Background Radiation frame, which may be considered as approximately fixed in the celestial sphere. For this and technical reasons, many authors have chosen the simpler situation (1) to explore some of the qualitative (and possible quantitative) effects of the noncommutative extensions. Indeed, the constant matrix $\theta^{\mu \nu}$ in (1) defines a bilinear skew-symmetric form $\Omega_{\theta} (a,b) = \sum_{\mu, \nu} a_{\mu} \theta^{\mu \nu} b_{\nu}$. If the dimensionality is even and this form is non-degenerate (this will be the case in the present work), then by a linear version of Darboux's Theorem \cite{Cannas} there exists a linear transformation which takes $\Omega_{\theta}$ into the standard symplectic form of classical mechanics (see below). This simplifies dramatically the approach to such systems. For these reasons, in this work we shall concentrate on the canonical case (1), and leave other extensions (such as (2)) for future a future work.

Adopting this point of view many authors have addressed the problem of testing the existence of space-time noncommutativity in nature by resorting to a quantum mechanical approximation to the complete quantum field theory. This theory of quantum mechanics with noncommutativity of the canonical type (1) is conventionally known as noncommutative quantum mechanics \cite{Mezincescu}-\cite{Gamboa}. This sort of theory also appears in the context of quantum constrained systems \cite{Girotti}.

Ho and Kao \cite{Ho} derived quantum mechanics from noncommutative quantum field theory in the nonrelativistic limit and showed that particles with opposite charges display opposite noncommutativity. Bolonek and Kosi\'nsky \cite{Bolonek} proved that Heisenberg's uncertainty principle is somewhat more restrictive in noncommutative quantum mechanics in the sense that e.g. on the noncommutative plane there are no quantum states saturating simultaneously more than one of the uncertainty relations. Most of the research thus far aims at finding upper bounds on the noncommutative parameters or at evaluating noncommutative corrections to ordinary quantum mechanical systems such as the harmonic oscillator, the Landau problem, the hydrogen atom, Lamb shift etc. \cite{Bertolami}-\cite{Gamboa}. Unfortunately, all these corrections seem to be too small to be detected experimentally with present technology.

In this work we shall consider the plane with spatial noncommutativity:
\begin{equation}
\left[\hat q_i, \hat q_j \right] = i \theta \epsilon_{ij} , \hspace{0.5 cm} \left[\hat q_i, \hat p_j \right] = i \hbar \delta_{ij} , \hspace{0.5 cm} \left[\hat p_i, \hat p_j \right] = 0 ,  \hspace{0.5 cm} i,j= 1,2,
\end{equation}
where $\epsilon_{12} = - \epsilon_{21} =1$, $\epsilon_{11}= \epsilon_{22}=0$ and $\theta$ is a real constant. We shall call this the extended Heisenberg algebra. All the particles appearing in the ensuing analysis will be assumed to display the same noncommutative parameter $\theta$. We shall frequently use the notation:
\begin{equation}
a \wedge b = a^T {\bf E} b= \sum_{i,j=1,2} \epsilon_{ij} a_i b_j = a_1 b_2 - a_2 b_1.
\end{equation}
Various arguments lead to the following estimate \cite{Carroll}:
\begin{equation}
\theta \leq 4 \times 10^{-40} m^2.
\end{equation}
To derive the noncommutative version of a quantum mechanical system one may follow two equivalent strategies. In the first case, one considers the usual Hamiltonian where now position dependent functions are multiplied by resorting to the Moyal $\star_{\theta}$-product \cite{Mezincescu}. If one considers e.g. a Hamiltonian of the form
\begin{equation}
\hat H = \frac{\hat p^2}{2m} +  V (\hat q),
\end{equation}
where $\hat q = (\hat q_1, \hat q_2 )$, $\hat p^2 = \hat p_1^2 + \hat p_2^2$, then the eigenvalue equation reads:
\begin{equation}
- \frac{\hbar^2}{2m} \nabla_q^2 \psi (q) + V(q) \star_{\theta} \psi (q) = E \psi (q),
\end{equation}
where $\nabla_q = \left( \partial / \partial q_1, \partial / \partial q_2 \right)$ and
\begin{equation}
A(q) \star_{\theta} B (q) = \left. \exp \left( \frac{i \theta}{2} \nabla_q \wedge \nabla_{q'}   \right) A(q) B (q') \right|_{q'=q}.
\end{equation}
One may however follow an alternative (but equivalent) route by noting that through a sort of "Seiberg-Witten map" \cite{Rosenbaum}
\begin{equation}
\hat R_i = \hat q_i + \frac{\theta}{2 \hbar} \epsilon_{ij} \hat p_j, \hspace{0.5 cm} \hat \Pi_i = \hat p_i,
\end{equation}
the noncommutative algebra (3) may be transformed into the usual Heisenberg algebra:
\begin{equation}
\left[ \hat R_i , \hat R_j \right] = \left[ \hat \Pi_i , \hat \Pi_j \right]=0 , \hspace{0.5 cm} \left[ \hat R_i , \hat \Pi_j \right] = i \hbar \delta_{ij} .
\end{equation}
In eq.(9) and henceforth we shall adopt the Einstein convention. Once the Hamiltonian (6) has been expressed in terms of the canonical variables $(\hat R, \hat \Pi)$, the usual quantization procedure follows. In particular, the space of states of noncommutative quantum mechanics remains $L^2 \left( \R^2, dR \right)$.

The aim of this research project is to study the emergence of ordinary quantum mechanics in the context of noncommutative quantum mechanics. We shall follow \cite{Hu3} and call this the noncommutative-commutative (NC-C) transition. This question is as pertinent as the question of how the classical world emerges in quantum theory \cite{Zeh}. It is well known that, in this case, there is no simple answer. Depending on the system \cite{Hu3} the criteria for classicality may vary. One may consider the formal limits $\hbar \to 0$, $n \to 0$ ($n$ is some quantum number), $N \to \infty$ ($N$ is the number of particles), or, alternatively, various approximations such as Ehrenfest's theorem, WKB approximations, high-temperature expansions or coherent states. In many cases one even resorts to combinations of various of these criteria.

Our strategy to induce a NC-C transition consists of coupling our noncommutative system to an external bath and thus treat it as an open system. This is well known in the context of environment-induced decoherence where a Brownian particle interacts with a heat bath at thermal equilibrium \cite{Zeh}. To keep our calculations as simple as possible we shall assume as in \cite{Caldeira} that the reservoir is constituted of identical bosonic particles interacting through a potential which has an absolute minimum and that there are no appreciable deviations from the equilibrium. Moreover we require that the coupling be weak so that only the linear response will be considered. This is commonly known as the {\it harmonic approximation}. The difference here is that all the particles involved live on the noncommutative plane with identical noncommutative parameter $\theta$. In the present paper we shall obtain an exact master equation (to all orders in $\theta$) for the reduced Wigner function of the Brownian particle which is the noncommutative extension of the Hu-Paz-Zhang equation \cite{Hu1}, \cite{Hu2}. In our derivation we shall follow closely the method of Halliwell and Yu \cite{Halliwell}. For this purpose a Weyl-Wigner formulation for noncommutative quantum mechanics is required \cite{Dias1}-\cite{Zachos4}. Some work in this direction has already been developed \cite{Rosenbaum}, \cite{Hu3}, \cite{Jing}. For our purposes we just need to recapitulate a few basic facts. In addition to the Moyal $\star_{\theta}$-product defined in (8) which involves only the position variables we shall also consider the usual Moyal $\star_{\hbar}$-product:
\begin{equation}
A(z) \star_{\hbar} B (z) = \left. \exp \left( \frac{i \hbar}{2} \nabla_z  \cdot {\bf J} \nabla_{z'} \right) A (z) B (z') \right|_{z'=z},
\end{equation}
where $\nabla_z = \left( \partial / \partial q, \partial / \partial p \right)$, $z= (q,p)$, etc and ${\bf J}$ is the $2d \times 2d$ symplectic matrix:
\begin{equation}
{\bf J} = \left(
\begin{array}{c r}
{\bf O}_{d \times d} &  {\bf I}_{d \times d}\\
- {\bf I}_{d \times d} & {\bf O}_{d \times d}
\end{array}
\right)
\end{equation}
and $d$ is the number of degrees of freedom (in this paper $d=2$). However, the basic algebraic structures for noncommutative quantum mechanics in phase space are the extended $\star$-product $(d=2)$:
\begin{equation}
A(z) \star B (z) = A(z) \star_{\hbar} \star_{\theta} B (z)= \left. \exp \left( \frac{i \hbar}{2} \nabla_z  \cdot {\bf J} \nabla_{z'} + \frac{i \theta}{2} \nabla_q \wedge \nabla_{q'} \right) A (z) B (z') \right|_{z'=z},
\end{equation}
and the extended Moyal bracket:
\begin{equation}
\left[A(z) , B (z) \right] = \frac{1}{i \hbar} \left( A (z) \star B (z) - B(z) \star A (z) \right).
\end{equation}
These expressions were derived in refs.\cite{Hu3,Jing,Bastos} using various methods. The dynamics of the noncommutative Wigner function is dictated by the extended Moyal equation:
\begin{equation}
\frac{\partial W}{\partial t}(z,t) = \left[H (z) , W (z,t) \right].
\end{equation}
In the previous equation, the noncommutative Wigner function $W(z)$ describes the state of the system. If the state is pure, then (for $d=2$) \cite{Jing}, \cite{Bastos}:
\begin{equation}
W(q,p) = \frac{1}{(\pi \hbar)^2} \int_{\bkR^2} dy ~ e^{- \frac{2i }{\hbar} y \cdot p} \psi (q-y)  \star_{\theta} \overline{\psi (q+y)},
\end{equation}
where the $\star_{\theta}$ product, which acts only on $q$, is given by (8). As usual, mixed states are described by convex combinations of states of the form (16). An important thing to remark is that, in general, these noncommutative Wigner functions are not the ordinary Wigner functions of quantum mechanics. Indeed, if we compute the marginal distribution
\begin{equation}
\int_{\bkR^2} d p ~ W(q,p)= \psi (q) \star_{\theta} \overline{\psi (q)},
\end{equation}
we conclude that it may not be everywhere non-negative. And thus, it cannot be interpreted as a true joint probability measure for $q_1, q_2$. Clearly, this is a consequence of the noncommutativity (3) and the associated uncertainty relations. Nevertheless, the existence of functions which are simultaneously noncommutative and ordinary Wigner functions is not precluded \cite{Bastos1}. In this work, we shall use as a criterion for the NC-C transition, the fact that the quasi-distribution describing the state of the Brownian particle is both a noncommutative and an ordinary Wigner function for a sufficiently long time (typically the relaxation time scale on which the particle reaches equilibrium with its environment).

The noncommutative Wigner function for a state at equilibrium $(\beta = \frac{1}{k_B T})$ with density matrix
\begin{equation}
\hat{\rho} = \left( Tr e^{- \beta \hat H} \right)^{-1} e^{- \beta \hat H}
\end{equation}
is given by \cite{Jing}
\begin{equation}
W(z) = N e_{\star}^{- \beta H},
\end{equation}
where $N$ is a normalization constant, $z$ may this time represent a collection of positions and momenta for several particles and $\phi (\beta , z) \equiv e_{\star}^{- \beta H}$ is the noncommutative exponential solution of the equation (regarded as a partial differential equation):
\begin{equation}
\frac{\partial \phi}{\partial \beta} = - \frac{1}{2} \left( H \star \phi + \phi \star H \right)
\end{equation}
and subject to the boundary condition:
\begin{equation}
\phi \left( \beta =0 , z \right) =1, \hspace{0.5 cm} \forall z \in T^*M.
\end{equation}
This is an extension of the formula obtained for the deformation quantization of ordinary quantum mechanics in \cite{Manogue,Hillery}. In these references the extended $\star$-product appearing e.g. in eq.(20) is replaced by the Moyal product $\star_{\hbar}$ (11).

\noindent
It is also important to note that the quantity
\begin{equation}
p [W] \equiv \int dz \hspace{0.2 cm} W^2 (z),
\end{equation}
associated with the state $W$ is still a measure of the purity of the system. In ref.\cite{Jing} the authors proved that there is an upper bound for the purity just like in the commutative case. Indeed if the system is in a state represented by the noncommutative Wigner function $W(q,p)$ then:
\begin{equation}
\left\{
\begin{array}{l l}
p[W] = \frac{1}{(2 \pi \hbar)^d}, & \mbox{if $W (z)$ is a pure state}\\
\\
p[W] < \frac{1}{(2 \pi \hbar)^d}, & \mbox{if $W (z)$ is a mixed state}
\end{array}
\right.
\end{equation}
This is all that will be required for the purposes of this paper in terms of the Weyl-Wigner formulation of noncommutative quantum systems. We have addressed this issue in more depth elsewhere \cite{Bastos}.

Before we conclude, a few remarks are in order. It may seem unnatural to have spatial noncommutativity and not the full phase space noncommutativity with the last commutation relation in (3) replaced by \cite{Bertolami}, \cite{Djemai}:
\begin{equation}
\left[\hat p_i, \hat p_j \right]= i \eta \epsilon_{ij}, \hspace{0.5 cm} i,j= 1,2.
\end{equation}
We shall however consider the simpler version in (3). There are various reasons for this. (i) Most authors \cite{Mezincescu}, \cite{Gamboa}, \cite{Dunne} regard the algebra in (3)  as the basis for noncommutative quantum mechanics. This is fact a direct prescription stemming from string theory, where the momentum noncommutativity is absent. (ii) Moreover, there are some interesting connections with the Landau problem and the quantum Hall effect \cite{Gamboa}, \cite{Polychronakos}, \cite{Susskind}. (iii) For the particular physical situation of the present work, the derivation of the master equation is cumbersome and an additional deformation of the momentum commutation relations would make this task even harder, while undermining our main objective: that of showing that the dissipative interaction of the system with an external environment is a suitable mechanism to induce the appearance of ordinary quantum mechanics in the realm of noncommutative quantum mechanics (we shall address this issue in section 6). (iv) As we mentioned before, in order to assess whether the NC-C transition has occurred, we need to know whether the state of the Brownian particle is described by a quasi-distribution which is simultaneously a noncommutative and an ordinary Wigner function. While it is by now firmly established what is meant by a noncommutative Wigner function (cf.(16)), when there is only spatial noncommutativity, there is no undisputed counterpart when the momenta are also noncommuting. For these reasons, we shall stick to the algebra (3).

Finally, let us summarize our motivation for writing this paper. Our main purpose is to derive a master equation governing the behaviour of a Brownian particle in interaction with an external environment in the context of a noncommutative extension of quantum mechanics. This equation provides an interesting starting point to address various (conceptual and structural) open issues on noncommutative quantum mechanics, namely:

\noindent
(i)
Can we understand under which conditions is a noncommutative system accurately described by the rules of ordinary quantum mechanics? This is an important point: if our universe is noncommutative and noncommutative quantum field theories are thus more fundamental than ordinary quantum theory, then we should be able to explain how ordinary quantum mechanics emerges in this context. The lessons learned in the past decade about the transition from ordinary quantum mechanics to classical mechanics will hopefully be a useful guide.  One approach that has been intensively explored in this context is that of dechoerence. Can a similar mechanism of dechoerence explain why our world looks "commutative" just as it partially explains why it looks mainly classical?
In this paper we prove that this seems to be the case. In fact the example studied in section 6 suggests that, in agreement with the claims of \cite{Hu3}, the noncommutative-commutative transition takes place before the quantum to classical transition. Moreover, the time scale of the former transition is (at least) 7 orders of magnitude shorter than that of the latter. And both of them are extremely short. This highlights the fact that we see a classical world and with some effort a quantum world, but not a "noncommutative" quantum world.

\noindent
(ii) Another important aspect concerns the symmetries of the theory. Many theoretical and experimental works (see e.g. \cite{Kostelecky}-\cite{Lehnert}) have recently addressed the breakdown of Lorentz invariance. Noncommutative quantum field theories are known to break this symmetry \cite{Nekrasov}. The system considered in this work is non-relativistic. However, we may regard it as a simplified toy model for addressing this problem. Indeed a two-dimensional harmonic oscillator Hamiltonian commutes with the angular momentum $\hat L = \hat q_1 \hat p_2 - \hat q_2 \hat p_1$, which is the generator of rotations on the plane. Once the oscillator is coupled to an external environment, then its angular momentum ceases to be conserved, and the rate of change is then determined by the torque exerted by the environment. Nevertheless this effect is controllable, and if the coupling is sufficiently weak, then during a finite lapse of time the angular momentum is approximately conserved. If we now add the extra spatial noncommutativity (3), then the rotational invariance is automatically broken even if we switch off the coupling to the environment. In this noncommutative case, there are thus two contributions to the breakdown of rotational invariance: the torque of the external bath, and the noncommutative contribution. If the coupling is sufficiently weak, then during a certain lapse of time (say $t_1$), the external torque will have only a residual influence on the angular momentum of the Brownian particle. However, as mentioned in (i), the time scale $t_0$ for the noncommutative-commutative transition is extremely short. After this instant $t_0$, we expect to see the noncommutative contribution to the angular momentum balance equation gradually fading away. And thus during the time interval $\left[t_0, t_1 \right]$, a rough conservation of the angular momentum is to be expected. This would be an interesting toy model to explain why Lorentz invariance is such a (experimentally) robust symmetry (even if our universe is noncommutative). We hope to consider this problem in a future work.

This paper is organized as follows. In the next section we obtain an expression for the noncommutative Wigner distribution of a collection of noninteracting noncommutative harmonic oscillators at thermal equilibrium as well as various expectation values which will be useful for the sequel. In section 3 we derive the exact noncommutative Hu-Paz-Zhang equation. In section 4, we obtain the equation that governs the dynamics of the purity and state sufficient conditions for a strict decrease of this quantity. As particular cases we address the weak coupling limit in section 5, and a noncommutative version of the Caldeira-Leggett model \cite{Caldeira} in section 6. In the latter case, we obtain an estimate for the time scale of the NC-C transition. Finally, in section 7 we present our conclusions. The lengthy derivations of the coefficients of the master equation and of the equilibrium distribution of the bath are relegated to the Appendices.

\section{Equilibrium distribution and momenta for a heat bath of noncommutative harmonic oscillators}

The purpose of this section is to state useful results concerning the equilibrium noncommutative Wigner distribution for a bath of $N$ noncommutative harmonic oscillators at temperature $T= \frac{1}{k_B \beta}$. All the formulae presented here are derived explicitly in Appendix 1. We assume the oscillators to be noninteracting, and to have masses $(m_n)$ and frequencies $(\omega_n)$. The corresponding variables $\left(\hat q^{(n)} , \hat p^{(n)} \right)$ obey the noncommutative algebra:
\begin{equation}
\left[\hat q_i^{(n)} , \hat q_j^{(m)} \right] = i \theta \delta_{n,m} \epsilon_{ij} , \hspace{1 cm} \left[\hat q_i^{(n)} , \hat p_j^{(m)} \right] = i \hbar \delta_{n,m} \delta_{i,j} , \hspace{1 cm}
\left[\hat p_i^{(n)} , \hat p_j^{(m)} \right] = 0.
\end{equation}
The Hamiltonian is:
\begin{equation}
\hat H \left( \left\{ \hat q^{(n)} , \hat p^{(n)} \right\} \right) = \sum_{n=1}^N \hat H^{(n)}
\left(\hat q^{(n)} , \hat p^{(n)} \right), \hspace{0.5 cm} \hat H^{(n)} \left( \hat q^{(n)} , \hat p^{(n)} \right) = \frac{ \left(\hat p^{(n)} \right)^2}{2m_n} + \frac{1}{2} m_n \omega_n^2 \left(\hat q^{(n)} \right)^2.
\end{equation}
The following quantities will be useful:
\begin{equation}
L^{(n)} =  q^{(n)} \wedge p^{(n)}, \hspace{0.5 cm} \lambda_n = \frac{m_n \omega_n^2 \theta}{2 \hbar}, \hspace{0.5 cm} M_n = \frac{m_n}{1 + (\lambda_n / \omega_n)^2} , \hspace{0.5 cm} \Omega_n = \omega_n \sqrt{1 + (\lambda_n / \omega_n )^2}.
\end{equation}
As the oscillators are noninteracting, the noncommutative Wigner function factorizes:
\begin{equation}
\left\{
\begin{array}{l}
W^b \left(\left\{  q^{(n)} ,  p^{(n)} \right\} \right) = \prod_{n=1}^N W_n^b \left( q^{(n)} , p^{(n)} \right) , \\\\
W_n^b \left(q^{(n)},p^{(n)} \right)=  \frac{1}{(\pi \hbar)^2} \tanh \left[\frac{\hbar \beta}{2} (\Omega_n + \lambda_n) \right] \tanh \left[\frac{\hbar \beta}{2} (\Omega_n - \lambda_n) \right] \exp \left[ - a_n (\beta) \left(p^{(n)} \right)^2 - c_n ( \beta) \left(q^{(n)} \right)^2 - 2 b_n (\beta) L^{(n)} \right],
\end{array}
\right.
\end{equation}
where:
\begin{equation}
\left\{
\begin{array}{l}
a_n (\beta) =  \frac{ (\Omega_n + \lambda_n )^2}{2 \hbar M_n \Omega_n^3 } \tanh\left[\frac{\hbar \beta}{2} (\Omega_n - \lambda_n ) \right] +\frac{ (\Omega_n - \lambda_n )^2}{2 \hbar M_n \Omega_n^3 } \tanh \left[\frac{\hbar \beta}{2} (\Omega_n + \lambda_n ) \right]\\
\\
c_n(\beta) = \frac{ M_n \Omega_n}{2 \hbar} \left\{ \tanh \left[\frac{\hbar \beta}{2} (\Omega_n - \lambda_n )  \right] +  \tanh \left[ \frac{\hbar \beta}{2} (\Omega_n + \lambda_n ) \right] \right\}\\
\\
b_n(\beta) = \frac{ (\Omega_n + \lambda_n )}{2 \hbar \Omega_n} \tanh \left[\frac{\hbar \beta}{2} (\Omega_n - \lambda_n ) \right] - \frac{ (\Omega_n - \lambda_n )}{2 \hbar \Omega_n} \tanh \left[\frac{\hbar \beta}{2} (\Omega_n + \lambda_n ) \right]
\end{array}
\right.
\end{equation}
For the purposes of this work, we shall need the following expectation values:
\begin{equation}
\begin{array}{l}
< \hat q_i ^{(n)}> = < \hat p_i ^{(n)}> = 0\\
\\
< \frac{\hat q_i ^{(n)} \hat p_j ^{(m)} + \hat p_j ^{(m)} \hat q_i ^{(n)}}{2}> =  - \frac{\hbar \delta_{n,m} \epsilon_{ij}}{4 \Omega_n} \left\{ \left( \Omega_n + \lambda_n \right) \coth \left[\frac{\hbar \beta}{2} \left( \Omega_n + \lambda_n \right) \right]
- \left( \Omega_n - \lambda_n \right) \coth \left[\frac{\hbar \beta}{2} \left( \Omega_n - \lambda_n \right) \right] \right\}\\
\\
< \hat p_i ^{(n)} \hat p_j ^{(m)}> =   \frac{\hbar \delta_{n,m} \delta_{ij} M_n \Omega_n}{4} \left\{ \coth \left[\frac{\hbar \beta}{2} \left( \Omega_n + \lambda_n \right) \right]
+  \coth \left[\frac{\hbar \beta}{2} \left( \Omega_n - \lambda_n \right) \right] \right\}\\
\\
< \frac{\hat q_i ^{(n)} \hat q_j ^{(m)} + \hat q_j ^{(m)} \hat q_i ^{(n)}}{2}> =   \frac{\hbar \delta_{n,m} \delta_{ij}}{4 M_n \Omega_n^3} \left\{ \left( \Omega_n + \lambda_n \right)^2 \coth \left[\frac{\hbar \beta}{2} \left( \Omega_n + \lambda_n \right) \right]
+ \left( \Omega_n - \lambda_n \right)^2 \coth \left[\frac{\hbar \beta}{2} \left( \Omega_n - \lambda_n \right) \right] \right\}
\end{array}
\end{equation}

\section{The noncommutative Brownian particle}

In this section we shall consider the combined system of a noncommutative quantum Brownian particle and the heat bath of the previous section. The Brownian particle is an oscillator of mass M and bare frequency $\Omega$. Its coordinates and momenta are $(q,p)$. We assume the particle to be linearly coupled to the environment so that the total Hamiltonian reads:
\begin{equation}
\hat H = \frac{\hat p^2}{2M} + \frac{1}{2} M \Omega^2 \hat q^2 + \sum_{n=1}^N \hat H^{(n)} + \hat q \cdot \sum_{n=1}^N C_n \hat q^{(n)},
\end{equation}
where $\hat H^{(n)}$ is as in eq.(26). As usual \cite{Caldeira}, \cite{Halliwell} we assume that at time $t=0$ the system and environment are uncorrelated:
\begin{equation}
W_0 \left(q,p; \left\{ q^{(n)}, p^{(n)} \right\} \right) = W_0^S (q,p) \cdot W_0^b \left( \left\{ q^{(n)}, p^{(n)} \right\} \right),
\end{equation}
where $W_0^b$ is given by (27-29). The Wigner function of the combined ensemble satisfies the noncommutative Moyal equation (15):
\begin{equation}
\begin{array}{c}
i \hbar \frac{\partial W}{\partial t} = H \star W - W \star H \Longleftrightarrow \frac{\partial W}{\partial t} = - \frac{p}{M} \cdot \nabla_q W + M \Omega^2 q \cdot \nabla_p W + \sum_n \left( - \frac{p^{(n)}}{m_n} \cdot \nabla_{q^{(n)}} + m_n \omega_n^2 q^{(n)} \cdot \nabla_{p^{(n)}} \right) W + \\
\\
+ \sum_n C_n \left( q^{(n)} \cdot \nabla_p + q \cdot \nabla_{p^{(n)}} \right) W + \frac{\theta}{\hbar} \sum_n \left( m_n \omega_n^2 q^{(n)} + C_n q \right) \wedge \nabla_{q^{(n)}} W + \frac{\theta}{\hbar} \left( M \Omega^2 q + \sum_n C_n q^{(n)} \right) \wedge \nabla_q W.
\end{array}
\end{equation}
The reduced Wigner function of the Brownian particle is obtained from W by tracing out the environment's degrees of freedom:
\begin{equation}
W_r (q,p) \equiv \int \left[\prod_n d q^{(n)} d p^{(n)} \right] W \left( q, p , \left\{q^{(n)} , p^{(n)} \right\} \right).
\end{equation}
Let us now perform this integration in eq.(33). We shall assume that $W$ and its first derivatives vanish at infinity, so that there will be no phase-space surface contributions. The result is:
\begin{equation}
\begin{array}{c}
\frac{\partial W_r}{\partial t} = - \frac{p}{M} \cdot \nabla_q W_r + M \Omega^2 q \cdot \nabla_p W_r + \sum_n C_n \nabla_p \cdot \int \left[\prod_m d q^{(m)} d p^{(m)} \right] q^{(n)}  W \left( q, p , \left\{q^{(m)} , p^{(m)} \right\} \right)+ \\
\\
+ \frac{\theta}{\hbar}  M \Omega^2 q  \wedge \nabla_q W_r -  \frac{\theta}{\hbar} \sum_n  C_n   \nabla_q \wedge \int \left[\prod_m d q^{(m)} d p^{(m)} \right] q^{(n)} W \left( q, p , \left\{q^{(m)} , p^{(m)} \right\} \right) .
\end{array}
\end{equation}
So we have to evaluate the following quantities:
\begin{equation}
G (q,p) \equiv  \sum_n  C_n \int \left[\prod_m d q^{(m)} d p^{(m)} \right] q^{(n)} W \left( q, p , \left\{q^{(m)} , p^{(m)} \right\} \right).
\end{equation}
Following the same argument as in \cite{Halliwell} we conclude that:
\begin{equation}
G (q,p) = \left[{\bf A} (t) q + {\bf B}(t) p + {\bf C}(t) \nabla_q + {\bf D}(t) \nabla_p \right] W_r (q,p),
\end{equation}
where ${\bf A}$, ${\bf B}$, ${\bf C}$, ${\bf D}$ are, for the time being, arbitrary time dependent $2 \times 2$ matrices. We then get from (35):
\begin{equation}
\begin{array}{c}
\frac{\partial W_r}{\partial t} = - \frac{p}{M} \cdot \nabla_q W_r + M \Omega^2 q \cdot \nabla_p W_r + \left( \nabla_p W_r \right) \cdot {\bf A}(t) q + \nabla_p  \cdot \left( {\bf B}(t) p W_r \right) + \nabla_p  \cdot \left( {\bf C}(t) \nabla_q W_r \right) + \nabla_p  \cdot \left( {\bf D}(t) \nabla_p W_r \right) + \\
\\
+ \frac{\theta}{\hbar}  M \Omega^2 q  \wedge \nabla_q W_r -  \frac{\theta}{\hbar}  \nabla_q \wedge \left({\bf A}(t) q W_r \right) -  \frac{\theta}{\hbar}  \nabla_q \wedge \left({\bf B}(t) p W_r \right) -  \frac{\theta}{\hbar}  \nabla_q \wedge \left({\bf C}(t) \nabla_q W_r \right) -  \frac{\theta}{\hbar}  \nabla_q \wedge \left({\bf D}(t) \nabla_p W_r \right) .
\end{array}
\end{equation}
It is important to remark that, in the commutative limit $(\theta \to 0)$ and for quadratic Hamiltonians, there is no difference between the Moyal equation (15) and the classical Liouville equation. In that case the quantum effects manifest themselves through the bath's equilibrium distribution (28) (with $\theta=0$). This is where Planck's constant makes its appearance. At the level of the master equation these quantum effects will be encoded solely in the matrix coefficients ${\bf A}(t), {\bf B}(t), {\bf C}(t), {\bf D}(t)$. In contrast with this behaviour, in the noncommutative case $(\theta>0)$, the quantum noncommutative effects appear explicitly in the master equation (38) even for quadratic Hamiltonians.

We conclude this section by presenting an expression for the matrix coefficients ${\bf A}(t), {\bf B}(t), {\bf C}(t), {\bf D}(t)$ in terms of physical quantities which characterize the environment. All the relevant equations are derived in the lengthy Appendix 2.

As in the commutative case the environment is characterized by a dissipation $(\eta_{ij} (t))$ and a noise kernel $(\nu_{ij} (t))$. These read:
\begin{equation}
\eta_{ij} (t) = \frac{d}{dt} \int_0^{+ \infty} \frac{d \omega}{\omega} \left[ I^+ (\omega) \delta_{ij}  \cos (\omega t) + I^- (\omega) \epsilon_{ij} \sin (\omega t) \right],
\end{equation}
and
\begin{equation}
\nu_{ij} (t) = \int_0^{+ \infty} d \omega \hspace{0.2 cm} \coth \left( \frac{\hbar \beta \omega}{2} \right) \left[I^+ (\omega) \delta_{ij}  \cos (\omega t) + I^- (\omega) \epsilon_{ij} \sin (\omega t) \right].
\end{equation}
Here $I^{\pm} (\omega)$ are the spectral densities:
\begin{equation}
I^{\pm} (\omega) = \sum_n \frac{C_n^2 \omega^2 }{4 m_n \omega_n^2 \Omega_n }  \left[\delta (\omega - \Omega_n - \lambda_n ) \pm \delta (\omega - \Omega_n + \lambda_n ) \right],
\end{equation}
where $\lambda_n, \Omega_n$ are as in eq.(27).
The matrix coefficients of the master equation, will also depend on the elementary functions $u_{ij} (s)$, $v_{ij} (s)$, linearly independent solutions of the homogeneous integrodifferential equations \cite{Hu1}, \cite{Hu2}:
\begin{equation}
\ddot \Sigma_{ij} (s) + \Omega^2 \Sigma_{ij} (s) - \frac{\theta}{\hbar} M \Omega^2 \epsilon_{ik} \dot \Sigma_{kj} (s) + \frac{2}{M} \int_0^s d \tau \hspace{0.2 cm} \eta_{kl} (s - \tau) \left( \delta_{ik} - \frac{M \theta}{\hbar} \epsilon_{ik} \frac{d}{d \tau} \right) \Sigma_{lj} (\tau) = 0,
\end{equation}
with the boundary conditions:
\begin{equation}
\left\{
\begin{array}{l l}
u_{ij} (s=0) = \delta_{ij}, & u_{ij} (s=t) =0,\\
v_{ij} (s=0) =0, & v_{ij} (s=t) = \delta_{ij}
\end{array}
\right.
\end{equation}
We can subsequently construct the functions:
\begin{equation}
G^{(1)}_{ij} (s, \tau) = \left\{
\begin{array}{l l}
L_{ik} (s, \tau) L_{kj}'^{-1} (\tau, \tau), & \mbox{if } s> \tau\\
0 , & \mbox{otherwise}
\end{array}
\right., \hspace{1 cm}
G^{(2)}_{ij} (s, \tau) = \left\{
\begin{array}{l l}
L_{ik} (s, \tau) L_{kj}'^{-1} (\tau, \tau), & \mbox{if } \tau > s\\
0 , & \mbox{otherwise}
\end{array}
\right.,
\end{equation}
with
\begin{equation}
L_{ij} (s, \tau) = u_{ij} (s) - v_{ik} (s) \left(v (\tau)\right)_{kl}^{-1} u (\tau)_{lj} , \hspace{1 cm}
L_{ij}' (\tau, \tau)= \left. \frac{\partial }{\partial s} L_{ij} (s, \tau) \right|_{s= \tau}
\end{equation}
Functions $G^{(k)}_{ij} (s, \tau)$ $(k=1,2)$ are the Green functions of eq.(147) (Appendix 2) with boundary conditions:
\begin{equation}
\left\{
\begin{array}{l l}
G_{ij}^{(1)} (s,s^-) =0 , & \left. \frac{d}{ds} G_{ij}^{(1)} (s, \tau) \right|_{ \tau = s^-} = \delta_{ij},\\
& \\
G_{ij}^{(2)} (s,s^+) =0 , & \left. \frac{d}{ds} G_{ij}^{(2)} (s, \tau) \right|_{ \tau = s^+} = \delta_{ij},
\end{array}
\right.
\end{equation}
Here and henceforth, we shall tacitly assume that the inverse matrices of $\left(\dot v_{ij} (s) \right)$ and $\left(\dot u_{ij} (s) \right)$ exist. We are now in a position to present the expressions for all the coefficients appearing in the master equation. The result is:
\begin{equation}
\begin{array}{l l}
A_{ij} (t) = & 2 \int_0^t ds \hspace{0.2 cm} \eta_{kl} (t-s) \left( \delta_{ik} - \frac{M \theta}{\hbar} \epsilon_{ik} \frac{d}{ds} \right) \left[v_{lj} (s) - u_{la} (s) \left(\dot u (t)\right)^{-1}_{ab}  \dot v_{bj} (t) \right] \\
& \\
B_{ij} (t) = &  \frac{2}{M} \int_0^t ds \hspace{0.2 cm} \eta_{lk} (t-s) \left( \delta_{il} - \frac{M \theta}{\hbar} \epsilon_{il} \frac{d}{ds} \right) \left\{u_{ka} (s) \left(\dot u (t)\right)^{-1}_{aj} + \frac{\theta M}{\hbar} \epsilon_{rj} \left[v_{kr} (s) - u_{ka} (s) \left(\dot u (t)\right)^{-1}_{ab}  \dot v_{br} (t) \right]  \right\} \\
& \\
C_{ij} (t)  = & \frac{\hbar}{M} \int_0^t d \lambda \hspace{0.2 cm} \Lambda_{jk}^{(1)} (t, \lambda) \nu_{ik} (t - \lambda) + \frac{2 \hbar}{M^2} \int_0^t ds \int_s^t d \tau \int_0^t d \lambda \hspace{0.2 cm} \Lambda_{jk}^{(1)} (t, \lambda) \Lambda_{lr}^{(2)} (s, \tau) \eta_{il} (t-s) \nu_{rk} ( \tau - \lambda)\\
& \\
D_{ij} (t) =  & \hbar \int_0^t d \lambda \hspace{0.2 cm} \left[\frac{d}{dt} G_{jk}^{(1)} (t, \lambda) \right] \nu_{ik} (t- \lambda) + \frac{2 \hbar}{M} \int_0^t ds \int_s^t d \tau \int_0^t d \lambda \hspace{0.2 cm} \eta_{il} (t-s) \left[\frac{d}{dt} G_{jk}^{(1)} (t, \lambda)\right] \Lambda_{lr}^{(2)} (s, \tau ) \nu_{rk} (\tau - \lambda)
\end{array}
\end{equation}
where
\begin{equation}
\Lambda_{ij}^{(k)} (s, \tau ) =  \left(\delta_{il} - \frac{\theta}{\hbar} M \epsilon_{il} \frac{d}{ds} \right) G_{lj}^{(k)} (s, \tau) , \hspace{0.5 cm} k=1,2.
\end{equation}
Two remarks are now in order. First of all the integrals in the previous expressions should be regarded as improper in the sense that e.g $\int_0^t ds = \lim_{\epsilon \to 0^+} \int_0^{t + \epsilon} ~ds$. This is also required in the commutative case \cite{Halliwell}. Secondly, as in the commutative case, all the matrices in the previous equation are in principle completely determined once the dissipation and noise kernels $\eta_{ij} (t)$, $\nu_{ij} (t)$ are given. In particular, they do not depend on the initial conditions of the Brownian particle.

\noindent
The structure of equations (47) looks very complicated. However, in most applications, one considers simplified situations, such as the weak coupling limit (see section 5) or the Ohmic high temperature Caldeira-Leggett model (see section 6).

\section{Purity}

A common issue in the context of dissipative quantum mechanics is that of finding general conditions of dissipativity \cite{Lidar}, \cite{Lindblad}, i.e conditions under which the purity of a state is guaranteed to decrease. Using the noncommutative Hu-Paz-Zhang equation derived in section 3 we may now obtain a set of sufficient conditions for this "H-theorem" to hold for the reduced Wigner function $W_r$. After a simple calculation we obtain:
\begin{equation}
\dot p [W_r] = Tr \left( {\bf B} - \frac{\theta}{\hbar} {\bf E} {\bf A} \right) p [W_r] - \int dz \hspace{0.2 cm} \nabla_z W_r \cdot {\bf {\cal J}} \nabla_z W_r,
\end{equation}
where $z$ and $\nabla_z$ are as in eq.(11). The $2 \times 2$ matrices ${\bf A}$, ${\bf B}$, ${\bf C}$ and ${\bf D}$ are as in (47), ${\bf E}$ is the $2 \times 2$ matrix with entries $\epsilon_{ij}$ and ${\bf {\cal J}}$ is the $4 \times 4$ matrix:
\begin{equation}
{\bf {\cal J}} = \left[
\begin{array}{c c}
\frac{\theta}{\hbar} ({\bf C}^T {\bf E} - {\bf E} {\bf  C} ) & {\bf C}^T - \frac{\theta}{\hbar} {\bf E} {\bf D}\\
\\
{\bf C} + \frac{\theta}{\hbar} {\bf D}^T {\bf E}& {\bf D} + {\bf D}^T
\end{array}
\right]
\end{equation}
Since $p[W_r] \ge 0$, we then have the following set of sufficient conditions for strict "purity" decrease:

\vspace{0.3 cm}
\noindent
(i) the $2 \times 2$ matrix ${\bf B} - \frac{\theta}{\hbar} {\bf E} {\bf A}$ has nonpositive trace; and

\noindent
(ii) the $4 \times 4$ matrix ${\bf {\cal J}}$ is positive semidefinite.

\section{The weak coupling limit}

As a particular case we consider a weak coupling between the Brownian particle and the heat bath. We will determine the matrices (47) to order ${\cal O} \left( C_n^2 \right)$. As $\eta_{ij} (t)$, $\nu_{ij} (t)$ are of order ${\cal O} \left( C_n^2 \right)$, we conclude from (47) that we shall need the functions $u_{ij} (t)$, $v_{ij} (t)$ only to zero-th order:
\begin{equation}
\ddot \Sigma_{ij} (s) + \Omega^2 \Sigma_{ij} (s) - \frac{\theta}{\hbar} M \Omega^2 \epsilon_{ik} \dot \Sigma_{kj} (s) =0.
\end{equation}
The general solution is given by:
\begin{equation}
\Sigma_{ij}  (s) = m_{ik} \rho_{kj} (s) + n_{ik} \lambda_{kj} (s),
\end{equation}
where $m_{ij}$, $n_{ij}$ are real constants and:
\begin{equation}
\left\{
\begin{array}{l}
\lambda_{ij} (s) = \frac{1}{2 \Sigma} \sum_{\sigma = \pm} \left[\delta_{ij} (\Sigma - \sigma \lambda ) \cos \left( (\Sigma + \sigma \lambda ) s \right) + \sigma \epsilon_{ij} (\Sigma - \sigma \lambda ) \sin \left( (\Sigma + \sigma \lambda )s \right) \right]\\
\\
\rho_{ij} (s) = \frac{1}{2 \Sigma} \sum_{\sigma = \pm} \left[\delta_{ij} \sin \left( (\Sigma + \sigma \lambda )s \right) - \sigma \epsilon_{ij}  \cos \left( (\Sigma + \sigma \lambda )s \right) \right]
\end{array}
\right.
\end{equation}
with
\begin{equation}
\lambda = \frac{M \Omega^2 \theta}{2 \hbar} , \hspace{1 cm} \Sigma = \Omega \sqrt{1 + \left( \frac{\lambda}{\Omega} \right)^2}.
\end{equation}
Notice that:
\begin{equation}
\lambda_{ij} \left( s =0 \right) = \dot \rho_{ij} \left( s =0 \right) = \delta_{ij}, \hspace{1 cm} \dot \lambda_{ij} \left( s =0 \right) = \rho_{ij} \left( s =0 \right)=0.
\end{equation}
To compute $A_{ij} (t)$, $B_{ij}(t)$ we do not really need the functions $u_{ij} (s)$, $v_{ij} (s)$, but rather the alternative set
\begin{equation}
R_{ij} (s) = v_{ij} (s) - u_{ik} (s) \left(\dot u (t) \right)_{kl}^{-1} \dot v_{lj} (t), \hspace{1 cm} T_{ij} (s) = u_{ik} (s) \left(\dot u (t) \right)_{kj}^{-1},
\end{equation}
which are equally linearly independent solutions of (42) with initial conditions:
\begin{equation}
R_{ij} \left( s =t \right) = \dot T_{ij} \left( s =t \right) = \delta_{ij}, \hspace{1 cm} \dot R_{ij} \left( s =t \right) = T_{ij} \left( s =t \right)=0
\end{equation}
From (52,55,57) we obtain:
\begin{equation}
R_{ij} (s) = \lambda_{ij} (s-t), \hspace{1 cm} T_{ij} (s) = \rho_{ij} (s-t)
\end{equation}
We then get from (47):
\begin{equation}
\left\{
\begin{array}{l}
A_{ij} (t) = 2 \int_0^t ds \hspace{0.2 cm} \eta_{kl} (s)  \left( \delta_{ik} + \frac{M \theta}{\hbar} \epsilon_{ik} \frac{d}{ds} \right) \lambda_{jl} (s)\\
\\
B_{ij} (t) = \frac{2}{M}  \int_0^t ds \hspace{0.2 cm} \eta_{kl} (s)  \left( \delta_{ik} + \frac{M \theta}{\hbar} \epsilon_{ik} \frac{d}{ds} \right) \left[ \rho_{jl} (s) - \frac{\theta M}{\hbar} \epsilon_{jr} \lambda_{rl} (s) \right]
\end{array}
\right.
\end{equation}
In the commutative limit $(\lambda \to 0, \Sigma \to \Omega)$,
\begin{equation}
\lambda_{ij} (t) \to \delta_{ij} \cos (\Omega t), \hspace{0.5 cm} \rho_{ij} (t) \to \frac{\delta_{ij}}{\Omega} \sin (\Omega t), \hspace{0.5 cm} \eta_{ij} (t) \to \delta_{ij} \eta (t),
\end{equation}
where:
\begin{equation}
\eta (t) = - \sum_n \frac{C_n^2}{2 m_n \omega_n} \sin (\omega_n t)
\end{equation}
is the commutative dissipation kernel. We then have:
\begin{equation}
A_{ij} (t) \to  2 \delta_{ij} \int_0^t ds \hspace{0.2 cm} \eta (s)  \cos (\Omega s) , \hspace{1 cm}  B_{ij} (t) \to  \frac{2 \delta_{ij}}{M \omega_n}  \int_0^t ds \hspace{0.2 cm} \eta (s)  \sin (\Omega s)
\end{equation}
as expected \cite{Halliwell}. To compute $C_{ij} (t)$, we need $(s < \tau)$:
\begin{equation}
G_{ij}^{(1)} (s , \tau) = \left[u_{ik} (s) v_{kl} (\tau) - v_{ik} (s) u_{kl} (\tau) \right] \left[\dot u_{lr} (\tau) v_{rj} (\tau) - \dot v_{lr} (\tau) u_{rj} (\tau) \right]^{-1}.
\end{equation}
Again, this is a solution of (42) in the variable $s$ with conditions:
\begin{equation}
\left. G_{ij}^{(1)} (s, \tau ) \right|_{s= \tau} =0 , \hspace{1 cm} \left. \frac{d}{ds} G_{ij}^{(1)} (s, \tau ) \right|_{s= \tau} = \delta_{ij}.
\end{equation}
Therefore the solution is:
\begin{equation}
G_{ij}^{(1)} (s, \tau ) = \rho_{ij} (s - \tau).
\end{equation}
The same holds for $G_{ij}^{(2)} (s, \tau)$ with $s < \tau$. Notice that in eq.(47) the second term in the expression for $C_{ij} (t)$ is of order greater than ${\cal O} \left( C_n^2 \right)$. We are thus left with:
\begin{equation}
C_{ij} (t) = \frac{ \hbar}{M} \int_0^t ds \hspace{0.2 cm} \nu_{ik} (s) \left(\delta_{jl} - \frac{\theta M}{\hbar} \epsilon_{jl} \frac{d}{d s} \right) \rho_{lk} (s).
\end{equation}
Similarly:
\begin{equation}
D_{ij} (t) = \hbar \int_0^t ds \hspace{0.2 cm} \nu_{ik} (s) \left(\delta_{jl} - \frac{\theta M}{\hbar} \epsilon_{jl} \frac{d}{d s} \right) \dot \rho_{lk} (s).
\end{equation}
In the commutative limit:
\begin{equation}
C_{ij} (t) \to  \frac{ \hbar \delta_{ij}}{M \Omega} \int_0^t ds \hspace{0.2 cm} \nu (s) \sin (\Omega s), \hspace{1 cm} D_{ij} (t) \to \hbar \delta_{ij} \int_0^t ds \hspace{0.2 cm} \nu (s) \cos (\Omega s),
\end{equation}
where
\begin{equation}
\nu (s) = \int_0^{+ \infty} d \omega \hspace{0.2 cm} \coth \left( \frac{\hbar \beta \omega}{2} \right) I (\omega) \cos (\omega t), \hspace{1 cm} I (\omega) = \sum_n \frac{C_n^2}{2 m_n \omega_n }\delta ( \omega - \omega_n )
\end{equation}
are the commutative noise kernel and spectral density, respectively. Again this is in agreement with known results.

\section{The Caldeira-Leggett model}

In this section we consider a simplified situation known as the Caldeira-Leggett model \cite{Caldeira}. We shall assume a continuum of oscillators with typical mass:
\begin{equation}
m_n \sim m , \hspace{1 cm} \forall n
\end{equation}
such that:
\begin{equation}
\sum_n C_n^2 \longrightarrow \int_0^{+ \infty} d \omega ~ \rho_D (\omega) C^2 (\omega),
\end{equation}
where
\begin{equation}
\rho_D (\omega) C^2 (\omega) = \left\{
\begin{array}{l l}
\frac{2m \eta  \omega^2}{\pi} & , \omega < \Lambda\\
0 & , \omega > \Lambda
\end{array}
\right.
\end{equation}
Here $\Lambda$ is a high-frequency cutoff, and $\eta$ is a damping constant. This model is Ohmic in the commutative limit in the sense that the spectral density (41) $I^+ (\omega)$ is proportional to $\omega$. In fact:
\begin{equation}
I^+ (\omega) = \frac{\eta \omega}{\pi}, \hspace{1 cm} I^- (\omega) =0  \hspace{1 cm} (\theta =0).
\end{equation}
We shall now make several approximations. We shall assume a high-temperature regime leading to a rapid localization, so that the localization timescale is much shorter than the relaxation timescale on which the system reaches equilibrium with its environment. Moreover we shall only consider noncommutative corrections to first order in $\theta$. This means that our description of the system will apply to short times and that $\sqrt{\theta}$ is small when compared with the typical coherence length. Altogether:
\begin{equation}
\frac{M \Lambda \theta}{\hbar} <<1, \hspace{1 cm} \frac{\theta \eta}{\hbar} << 1, \hspace{1 cm} \frac{\hbar \Lambda}{k_B T} << 1.
\end{equation}
Finally, we consider the typical mass of the bath to be much smaller than that of the Brownian particle:
\begin{equation}
\frac{m}{M} << 1.
\end{equation}
Under these assumptions and after some calculation, we obtain from (39,40):
\begin{equation}
\left\{
\begin{array}{l}
\eta_{ij} (t) \sim \eta \left( \delta_{ij} + \frac{m \theta}{2 \hbar} \epsilon_{ij} \frac{d}{dt} \right) \delta' (t)\\
\\
\nu_{ij} (t) \sim \frac{2 \eta k_B T}{\hbar} \left( \delta_{ij} + \frac{m \theta}{2 \hbar} \epsilon_{ij} \frac{d}{dt} \right) \delta (t)
\end{array}
\right.
\end{equation}
where as usual, we used the representation of the delta function:
\begin{equation}
\delta (t) = \lim_{\Lambda \to + \infty} \frac{\sin (\Lambda t)}{\pi t}.
\end{equation}
Here and henceforth, the symbol $\sim$ denotes the approximation (74,75).

To compute the coefficients ${\bf A},{\bf B},{\bf C},{\bf D}$ of the master equation, we need the functions $R_{ij} (s), T_{ij} (s)$ as they were defined in (56). In fact, we shall only need to know their boundary conditions (57), and their differential equation (42) to order zero in $\theta$. From (47,57,76), we obtain after some integrations by parts:
\begin{equation}
A_{ij} (t) \sim - \frac{2 \eta M \theta}{\hbar} \epsilon_{il} \ddot{R}_{lj}^{(0)} (t) - 2 \eta \delta_{ij} \delta (0) - \frac{\eta m \theta}{\hbar} \delta' (0) \epsilon_{ij},
\end{equation}
where $R_{ij}^{(0)} (s)$ is the zeroth-order (in $\theta$) approximation of $R_{ij} (s)$. To this order, we have from eq.(42):
\begin{equation}
\ddot{R}_{ij}^{(0)} (t) = - \Omega_{ren}^2 R_{ij}^{(0)} (t) - \frac{2 \eta}{M} \dot R_{ij}^{(0)} (t) = - \Omega_{ren}^2 \delta_{ij}.
\end{equation}
Here $\Omega_{ren}$ is the renormalized frequency \cite{Caldeira}, \cite{Halliwell}:
\begin{equation}
\Omega_{ren}^2 = \Omega^2 - \frac{2 \eta}{M} \delta (0).
\end{equation}
Altogether:
\begin{equation}
A_{ij} (t) \sim  - 2 \eta \delta_{ij} \delta (0) + \frac{2 \theta \eta M}{\hbar} \Omega_{ren}^2 \epsilon_{ij}.
\end{equation}
Notice that, as in the commutative case, $A_{ij} (t)$ is time-independent. In a similar fashion, we obtain:
\begin{equation}
B_{ij} (t) \sim   \frac{2 \eta}{M} \delta_{ij}  + \frac{4 \theta \eta^2}{M \hbar}  \epsilon_{ij},
\end{equation}
where we used
\begin{equation}
\ddot T_{ij}^{(0)} (t) = - \frac{2 \eta}{M} \delta_{ij}.
\end{equation}
Similarly, to compute $C_{ij} (t)$, we perform a few integrations by parts and use the boundary conditions (46) for the Green functions $G_{ij}^{(1)} (s, \tau)$, $G_{ij}^{(2)} (s, \tau)$. The result reads:
\begin{equation}
C_{ij} (t) \sim 2 \eta K_B T \frac{\theta}{\hbar} \epsilon_{ij}.
\end{equation}
Notice that this term is absent in the commutative limit.

Finally, to compute $D_{ij} (t)$, we need $\left. \frac{d^2}{dt^2} G_{ij}^{(1)} (t, \tau) \right|_{\tau = t}$ to zero-th order in $\theta$. This can be evaluated from eq.(147) (Appendix 2):
\begin{equation}
\left. \frac{d^2}{dt^2} G_{ij}^{(1)} (t, \tau) \right|_{\tau = t} = \frac{2 \eta}{M} \delta_{ij} \hspace{1 cm} (\theta =0).
\end{equation}
We then obtain:
\begin{equation}
D_{ij} (t) \sim 2 \eta k_B T \delta_{ij} + 2 \eta k_B T \frac{\theta}{\hbar} \sigma \epsilon_{ij}, \hspace{1 cm} \sigma \equiv 2 \eta - \frac{m}{2} \delta (0).
\end{equation}
As in the commutative limit, in the Caldeira-Leggett model all the coefficients in the master equation are constant. Upon substitution of these coeficients in the master equation (38), we obtain:
\begin{equation}
\begin{array}{c}
\frac{\partial W_r}{\partial t} \sim - \frac{p}{M} \cdot \nabla_q W_r + M \Omega_{ren}^2 q  \cdot \nabla_p W_r + \frac{2 \eta}{M} \nabla_p \cdot \left(p  W_r \right) +  2\eta k_B T
\nabla_p^2 W_r \\
\\
- \frac{2 \theta \eta M \Omega_{ren}^2}{\hbar} q \wedge \nabla_p W_r +  \frac{4 \theta \eta^2}{\hbar M} \nabla_p \wedge \left(p W_r \right) - \frac{4 \theta \eta k_B T}{\hbar} \nabla_q \wedge \nabla_p W_r + \\
\\
+ \frac{\theta M \Omega_{ren}^2}{\hbar} q \wedge \nabla_q  W_r -  \frac{2 \theta \eta}{\hbar M} \nabla_q \wedge \left(p W_r \right)
\end{array}
\end{equation}
For this model, we obtain from (87) or (49), the following equation for the purity of the system:
\begin{equation}
\dot p \left[W_r \right] = \frac{ 4 \eta}{M} p \left[W_r \right] - 4 \eta k_B T \int dq \int dp ~ \left| \nabla_p W_r (q,p) \right|^2.
\end{equation}
We conclude that the purity obeys the same dynamical equation as in the commutative limit. To observe differences in this quantity, we should consider higher order corrections in our approximation.

We now solve the master equation (87), but not in its full generality. We shall set $\Omega_{ren}=0$ and, in the spirit of (74), we shall take the limit $T \to + \infty$, $\eta \to 0$, while keeping the product:
\begin{equation}
\Gamma \equiv 2 \eta k_B T
\end{equation}
constant. Altogether, we obtain:
\begin{equation}
\frac{\partial W_r}{\partial t} \sim - \frac{p}{M} \cdot \nabla_q W_r  + \Gamma
\nabla_p^2 W_r  - \frac{2 \theta \Gamma}{\hbar} \nabla_q \wedge \nabla_p W_r
\end{equation}
Strictly speaking this is not a Fokker-Planck equation, as the diffusion matrix is not positive defined.

Similarly, in eq.(88) we obtain:
\begin{equation}
\dot p \left[W_r \right] =  - 2 \Gamma  \int dq \int dp ~ \left| \nabla_p W_r (q,p) \right|^2.
\end{equation}
The purity is thus strictly decreasing.

Using standard methods \cite{Brodier}, \cite{Zeh}, we obtain the solution:
\begin{equation}
W_r (q,p,t) \sim \int d q' \int dp' ~ G_t \left(q - \frac{t}{M} p - q' , p-p' \right) W_0 (q',p'),
\end{equation}
where $W_0 (q,p)$ is the initial noncommutative Wigner quasi-distribution:
\begin{equation}
W_r (q,p, t=0) = W_0 (q,p),
\end{equation}
and $G_t (q,p)$ is the Gaussian kernel:
\begin{equation}
G_t (z) \sim 3 \left( \frac{M}{2 \pi \Gamma t^2} \right)^2 \exp \left( - z^T {\bf M}_t z \right) , \hspace{1 cm} z^T =(q,p).
\end{equation}
The matrix ${\bf M}_t$ reads:
\begin{equation}
{\bf M}_t=
\frac{3 M^2 }{\Gamma t^3} \left(
\begin{array}{l r}
{\bf I}_{2 \times 2} & \frac{t}{M} {\bf I}_{2 \times 2} + \frac{\theta}{\hbar} {\bf E}\\
& \\
 \frac{t}{M} {\bf I}_{2 \times 2} - \frac{\theta}{\hbar} {\bf E} &  \frac{t^2}{3M^2} {\bf I}_{2 \times 2}
\end{array}
\right)
\end{equation}
So $W_r (q,p,t)$ is the convolution (modulo a symplectic transformation) of a noncommutative Wigner function with a Gaussian. Once the Gaussian satisfies the Heisenberg uncertainty relations
\begin{equation}
\Delta q_1 \Delta q_2 \sim \frac{\theta}{2}
\end{equation}
this convolution will become a Wigner quasi-distribution \cite{Bastos1}. As mentioned in the introduction, we assume that for the NC-C transition to occur, then a necessary (possibly not sufficient) condition should be the fact that the noncommutative Wigner function describing the state of the system become equally an ordinary Wigner function. This seems to be a reasonable criterion in analogy with the transition from ordinary quantum mechanics to classical mechanics. Indeed, in this transition the density matrix of the Brownian particle becomes approximately diagonal in the position basis. A necessary (but not sufficient) condition for this is that the corresponding Wigner function be positive. Notice also that if the noncommutative Wigner function becomes equally an ordinary Wigner function, then in particular, the marginal distribution (17), will be everywhere non-negative, regardless of whether it was positive or not at the initial time. This will allow the usual quantum mechanical probabilistic interpretation for the position variables compatible with a course graining resolving areas greater than $\theta$.

If we compute $\Delta q_i$ for the Gaussian $G_t (q,p)$, we obtain:
\begin{equation}
\Delta q_i \sim \sqrt{\frac{2 \Gamma t^3}{3 M^2}}.
\end{equation}
From (96,97), we thus get:
\begin{equation}
t_{\theta} \sim \sqrt[3]{\frac{3 M^2 \theta}{4 \Gamma}}.
\end{equation}
To this order in $\theta$ the noncommutative Wigner function becomes a Wigner function at $t_{\theta}$ irrespective of its initial distribution $W_0 (q,p)$. Strictly speaking, this may happen even before $t_{\theta}$. But what our analysis shows is that from $t_{\theta}$ onwards, it will certainly be a Wigner function. This situation is analogous to that described in \cite{Diosi1}, \cite{Brodier}.

It is instructive to compare this time scale with the typical time scale of decoherence, where the transition from ordinary quantum mechanics to classical mechanics takes place:
\begin{equation}
t_D \simeq \sqrt{\frac{\hbar M}{\Gamma}}.
\end{equation}
From (98,99), we obtain:
\begin{equation}
t_{\theta} \sim \sqrt[3]{\frac{M \theta t_D^2}{\hbar}}.
\end{equation}
Using the estimate (5), we get
\begin{equation}
t_{\theta} < 10^{-2} \sqrt[3]{Mt_D^2},
\end{equation}
if $M$ is expressed in kg and $t_{\theta}, t_D$ in seconds. As an example consider an electron in a medium constituted of air molecules at $T=300 K$ and pressure of 1 atm. For this situation we have \cite{Zeh}:
\begin{equation}
\Gamma \sim 10^{-33} m^2 kg^2 s^{-3}.
\end{equation}
We thus obtain:
\begin{equation}
t_D \sim 10^{-15} s \hspace{1 cm} t_{\theta} \sim 10^{-22} s,
\end{equation}
were we used the estimate (5) for $\theta$. And thus the time scale for the NC-C transition is much shorter than $t_D$. And one should keep in mind that (5) is an upper bound. Therefore, $t_{\theta}$ should be even smaller than our estimate (103). Notice that this result gives countenance to the claim in \cite{Hu3}, that the NC-C transition takes place before the transition from ordinary quantum mechanics to classical mechanics.

\section{Conclusions and outlook}

Let us restate our results. We computed the equilibrium Wigner distribution for the reservoir of noncommutative harmonic oscillators as well as the first and second order moments of this distribution. Assuming uncorrelated but otherwise arbitrary initial distributions for the Brownian particle and the bath, we derived the exact noncommutative extension of the Hu-Paz-Zhang master equation to all orders in $\theta$. By construction our equation is Markovian and of the Lindblad form, which ensures the positivity of the evolution. Moreover, and contrary to what happens in the commutative limit, the quantum effects appear explicitly in the master equation (38) and not just at the level of the matrix coefficients ${\bf A}(t), {\bf B}(t), {\bf C}(t), {\bf D}(t)$. We therefore expect these noncommutative corrections to lead to qualitatively different results. We also stated sufficient conditions for strict decrease of purity of the Brownian particle. Finally, we considered the particular cases of the weak coupling limit and the Caldeira-Leggett model. In the latter case, we solved the master equation and obtained an estimate for the time scale of the NC-C transition. For an electron at room temperature and pressure it is at least seven orders of magnitude shorter than the decoherence time scale. Just as the decoherence mechanism explains why universe looks mainly classical, it also seems to explain why it is mainly commutative. Our result supports the claim made in \cite{Hu3} that the NC-C transition takes place before the quantum-classical transition.

In a future work, we expect to further explore this master equation \cite{Bastos1}. In particular, we intend to study a coloured environment, entanglement, and investigate the appearance of certain symmetries once the NC-C transition occurs. We are obviously thinking of toy models for Lorentz invariance.
As there seem to be different time scales for the various transitions, we could therefore consider a situation where various systems (with different decohering time scales coexist). We may thus envisage the possibility of extending hybrid ensembles of coupled classical and quantum subsystems \cite{Diosi}-\cite{Dias4} to hybrid classical-quantum commutative-quantum noncommutative systems.

\appendix

\section*{Appendix 1}

Our aim is to derive the various formulae presented in section 2. Since the oscillators are noninteracting, we may start by evaluating the distribution for a single particle of mass $m$ and frequency $\omega$. Its position and momentum variables are $\hat q=(\hat q_1,\hat q_2)$, $\hat p=(\hat p_1,\hat p_2)$, respectively, and we assume that they obey the noncommutative algebra (3). The Hamiltonian reads:
\begin{equation}
\hat H \left( \hat q, \hat p \right) = \frac{\hat p^2}{2m} + \frac{1}{2} m \omega^2 \hat q^2.
\end{equation}
In terms of $\hat R$ and $\hat \Pi$ (cf.(9)), the Hamiltonian may be expressed as:
\begin{equation}
\hat H \left( \hat q, \hat p \right) =  \hat{{\cal H}} \left( \hat R, \hat \Pi \right) = \frac{1}{2m} \left[1 + \left( \frac{m \omega \theta}{2 \hbar} \right)^2 \right] \hat \Pi^2 + \frac{1}{2} m \omega^2 \hat R^2 - \frac{m \omega^2 \theta}{2 \hbar} \hat{{\cal L}},
\end{equation}
where $\hat{{\cal L}} \equiv \hat R \wedge \hat \Pi$ is the angular momentum. The equilibrium density matrix at temperature $T$ is then:
\begin{equation}
\hat{\rho} = {\cal N} \exp \left[- \beta \hat H \left( \hat q, \hat p \right) \right] =  {\cal N} \exp \left[- \beta \hat{{\cal H}} \left( \hat R, \hat \Pi \right) \right],
\end{equation}
where ${\cal N}$ is a normalization constant. To obtain the corresponding Wigner function in noncommutative phase-space we may resort to the generalized Weyl-Wigner map $W_{(R, \Pi)}^{(q,p)}$ introduced in \cite{Dias1}. Indeed, it is easy to prove that the application of this map induces in the phase space $T^*M$ the extended $\star$-product $(\star= \star_{\hbar} \star_{\theta})$ mentioned in the introduction \cite{Bastos}. This map consists of applying the Weyl-map based on the Heisenberg algebra $(R,\Pi)$ (cf(10)) and, subsequently, performing the phase-space version of the operator diffeomorphism (9). We are thus looking for the noncommutative exponential
\begin{equation}
\phi \left(\beta, R, \Pi \right) \equiv \exp_{\star_{\hbar}} \left\{ - \frac{\beta}{2m} \left[1 + \left( \frac{m \omega \theta}{2 \hbar} \right)^2 \right] \Pi^2  - \frac{\beta}{2} m \omega^2  R^2 + \frac{m \omega^2 \theta \beta}{2 \hbar} {\cal L} \right\},
\end{equation}
where $\star_{\hbar}$ is the usual Moyal product based on the Heisenberg variables $(R, \Pi)$ (cf.(11)). Let us rewrite the previous equation as:
\begin{equation}
\phi \left(\beta, R, \Pi \right) = \psi \left(k_1, k_2 , R, \Pi \right) = \exp_{\star_{\hbar}} \left[i k_1 \left( \frac{\Pi^2}{2M} + \frac{1}{2} M \Omega^2 R^2 \right) + i k_2 \Omega {\cal L} \right],
\end{equation}
where
\begin{equation}
\left\{
\begin{array}{l}
M = \frac{m}{1 + (\lambda / \omega)^2}, \hspace{0.5 cm} \Omega = \omega \sqrt{1 + (\lambda / \omega )^2}, \hspace{0.5 cm} \lambda = \frac{m \omega^2 \theta }{2 \hbar}\\
\\
k_1 = i \beta , \hspace{0.5 cm} k_2 = -i \beta \frac{\lambda}{\Omega}
\end{array}
\right.
\end{equation}
The exponential (108) is well known. We have computed it elsewhere \cite{Dias2} (see also \cite{Flato}) in the context of the two-dimensional commutative harmonic oscillator. The solution is:
\begin{equation}
\begin{array}{c}
\psi \left( k_1, k_2, R, \Pi \right) = \cos^{-1} \left[\frac{\hbar \Omega}{2} (k_1 + k_2) \right] \cos^{-1} \left[\frac{\hbar \Omega}{2} (k_1 - k_2) \right] \times \\
\\
\times \exp \left\{\frac{i}{\Omega \hbar} \tan \left[ \frac{\hbar \Omega}{2} (k_1+k_2) \right] \left( \frac{\Pi^2}{2 M} + \frac{1}{2} M \Omega^2 R^2 + \Omega {\cal L} \right) + \frac{i}{\Omega \hbar} \tan \left[ \frac{\hbar \Omega}{2} (k_1-k_2) \right] \left( \frac{\Pi^2}{2 M} + \frac{1}{2} M \Omega^2 R^2 - \Omega {\cal L} \right) \right\}.
\end{array}
\end{equation}
To finish our calculation we need to go back to the coordinates $(q,p)$ using (9). A simple calculation leads to:
\begin{equation}
\phi \left( \beta , q,p \right) = \cosh^{-1} \left[\frac{\hbar \beta}{2} (\Omega + \lambda ) \right] \cosh^{-1} \left[\frac{\hbar \beta}{2} (\Omega - \lambda ) \right] \times \exp \left[- a (\beta) p^2 - c(\beta) q^2 - 2 b (\beta) L \right],
\end{equation}
where this time $L = q \wedge p$ and
\begin{equation}
\left\{
\begin{array}{l}
a(\beta) =  \frac{ (\Omega + \lambda )^2}{2 \hbar M \Omega^3 } \tanh\left[\frac{\hbar \beta}{2} (\Omega - \lambda ) \right] +\frac{ (\Omega - \lambda )^2}{2 \hbar M \Omega^3 } \tanh \left[\frac{\hbar \beta}{2} (\Omega + \lambda ) \right]\\
\\
c(\beta) = \frac{ M \Omega}{2 \hbar} \left\{ \tanh \left[\frac{\hbar \beta}{2} (\Omega - \lambda )  \right] +  \tanh \left[ \frac{\hbar \beta}{2} (\Omega + \lambda ) \right] \right\}\\
\\
b(\beta) = \frac{ (\Omega + \lambda )}{2 \hbar \Omega} \tanh \left[\frac{\hbar \beta}{2} (\Omega - \lambda ) \right] - \frac{ (\Omega - \lambda )}{2 \hbar \Omega} \tanh \left[\frac{\hbar \beta}{2} (\Omega + \lambda ) \right]
\end{array}
\right.
\end{equation}
To check our result, we remark the following. The exponential (111) is formally a solution of (20,21) with $T^* M \simeq \R^4$. We have checked explicitly that our expression (111) is indeed a solution. Finally, from (19,111), we conclude that the equilibrium Wigner distribution is:
\begin{equation}
W (q,p) =  N (\beta) \exp \left[- a (\beta) p^2 - c (\beta) q^2 - 2 b (\beta) L \right],
\end{equation}
where $N(\beta)$ is the normalization constant. To derive $N( \beta )$ let us first define the following generating function:
\begin{equation}
{\cal Z} (\xi , \eta) \equiv \int d q \int d p \hspace{0.2 cm} W(q,p) \exp \left( \xi \cdot q + \eta \cdot p \right) = \frac{\pi^2 N}{ac - b^2} \exp \left[ \frac{-a \xi^2 - c \eta^2 - 2 b \eta \wedge \xi}{4 (b^2 - ac) } \right].
\end{equation}
The normalization is then given by the condition ${\cal Z} (0,0) =1$. We thus get:
\begin{equation}
N(\beta) = \frac{ac -b^2}{\pi^2} = \frac{1}{(\pi \hbar)^2} \tanh \left[\frac{\hbar \beta}{2} (\Omega + \lambda) \right] \tanh \left[\frac{\hbar \beta}{2} (\Omega - \lambda) \right].
\end{equation}
We finally have:
\begin{equation}
W (q,p)=  \frac{1}{(\pi \hbar)^2} \tanh \left[\frac{\hbar \beta}{2} (\Omega + \lambda) \right] \tanh \left[\frac{\hbar \beta}{2} (\Omega - \lambda) \right] \exp \left[- a (\beta) p^2 - c( \beta) q^2 - 2 b (\beta) L \right].
\end{equation}
Notice that in the commutative limit $\theta \to 0$, the following holds:
\begin{equation}
\left\{
\begin{array}{l l}
\Omega \to \omega, \hspace{1 cm} \lambda \to 0 , &  b ( \beta) \to 0\\
& \\
a (\beta) \to  \frac{1}{m \omega \hbar} \tanh \left(\frac{ \hbar \omega \beta}{2} \right), & c (\beta) \to  \frac{m \omega}{\hbar} \tanh \left( \frac{ \hbar \omega \beta}{2} \right)
\end{array}
\right.
\end{equation}
Consequently:
\begin{equation}
W(q,p)  \longrightarrow \left[ \frac{\tanh \left( \frac{\hbar \omega \beta}{2} \right) }{\pi \hbar} \right]^2 \exp \left[ - \frac{2}{\omega \hbar} \tanh \left( \frac{\hbar \omega \beta}{2} \right) \left( \frac{p^2}{2m} + \frac{1}{2} m \omega^2 q^2 \right) \right],
\end{equation}
which is the correct result for a two-dimensional commutative harmonic oscillator.

\vspace{0.3 cm}
\noindent
Let us now consider $N$ noninteracting harmonic oscillators with masses $m_n$ and frequencies $\omega_n$. The corresponding variables $\left(\hat q^{(n)}, \hat p^{(n)} \right)$ obey the noncommutative algebra (25). Our previous analysis reveals that the equilibrium noncommutative Wigner functions is given by eqs.(27-29). The corresponding generating function is:
\begin{equation}
{\cal Z}^b \left( \left\{ \xi^{(n)} , \eta^{(n)} \right\} \right) = \prod_{n=1}^N {\cal Z}_n^b \left(\xi^{(n)} , \eta^{(n)} \right) , \hspace{0.5 cm} {\cal Z}_n^b \left(\xi^{(n)} , \eta^{(n)} \right) = \exp \left[ \frac{- a_n  \left( \xi^{(n)} \right)^2 - c_n  \left( \eta^{(n)} \right)^2 -  2 b_n  \xi^{(n)} \wedge \eta^{(n)} }{4 \left( b_n^2 - a_n c_n \right)} \right].
\end{equation}
Let us now compute some expectation values using this generating function:
\begin{equation}
< \hat q_i ^{(n)}> = \left. \frac{\partial}{\partial \xi_i^{(n)}} {\cal Z}^b \left( \left\{ \xi^{(n)} , \eta^{(n)} \right\} \right) \right|_{\left( \left\{ \xi^{(n)} , \eta^{(n)} \right\} \right) =0} = \left. \frac{\partial}{\partial \xi_i^{(n)}} {\cal Z}_n^b \left( \xi^{(n)} , \eta^{(n)} \right) \right|_{\left(\xi^{(n)} , \eta^{(n)}\right) =0} =0.
\end{equation}
Similarly:
\begin{equation}
< \hat p_i ^{(n)}> = 0
\end{equation}
We also have
\begin{equation}
\begin{array}{c}
< \frac{\hat q_i ^{(n)} \hat p_j ^{(m)} + \hat p_j ^{(m)} \hat q_i ^{(n)}}{2}> = \left. \frac{\partial}{\partial \xi_i^{(n)}} \frac{\partial}{\partial \eta_j^{(m)}} {\cal Z}^b \left( \left\{ \xi^{(n)} , \eta^{(n)} \right\} \right) \right|_{\left( \left\{ \xi^{(n)} , \eta^{(n)} \right\} \right) =0} = \delta_{n,m} \left. \frac{\partial}{\partial \xi_i^{(n)}} \frac{\partial}{\partial \eta_j^{(n)}} {\cal Z}_n^b \left( \xi^{(n)} , \eta^{(n)} \right) \right|_{\left(  \xi^{(n)} , \eta^{(n)} \right) =0} = \\
\\
= - \frac{\hbar \delta_{n,m} \epsilon_{ij}}{4 \Omega_n} \left\{ \left( \Omega_n + \lambda_n \right) \coth \left[\frac{\hbar \beta}{2} \left( \Omega_n + \lambda_n \right) \right]
- \left( \Omega_n - \lambda_n \right) \coth \left[\frac{\hbar \beta}{2} \left( \Omega_n - \lambda_n \right) \right] \right\}
\end{array}
\end{equation}
and
\begin{equation}
\begin{array}{c}
< \hat p_i ^{(n)} \hat p_j ^{(m)}> =  \delta_{n,m} \left. \frac{\partial}{\partial \eta_i^{(n)}} \frac{\partial}{\partial \eta_j^{(n)}} {\cal Z}_n^b \left( \xi^{(n)} , \eta^{(n)} \right) \right|_{\left(  \xi^{(n)} , \eta^{(n)} \right) =0} = \\
\\
= \frac{\hbar \delta_{n,m} \delta_{ij} M_n \Omega_n}{4} \left\{ \coth \left[\frac{\hbar \beta}{2} \left( \Omega_n + \lambda_n \right) \right]
+  \coth \left[\frac{\hbar \beta}{2} \left( \Omega_n - \lambda_n \right) \right] \right\}
\end{array}
\end{equation}
Finally:
\begin{equation}
\begin{array}{c}
< \frac{\hat q_i ^{(n)} \hat q_j ^{(m)} + \hat q_j ^{(m)} \hat q_i ^{(n)}}{2}> =  \delta_{n,m} \left. \frac{\partial}{\partial \xi_i^{(n)}} \frac{\partial}{\partial \xi_j^{(n)}} {\cal Z}_n^b \left( \xi^{(n)} , \eta^{(n)} \right) \right|_{\left(  \xi^{(n)} , \eta^{(n)} \right) =0} = \\
\\
= \frac{\hbar \delta_{n,m} \delta_{ij}}{4 M_n \Omega_n^3} \left\{ \left( \Omega_n + \lambda_n \right)^2 \coth \left[\frac{\hbar \beta}{2} \left( \Omega_n + \lambda_n \right) \right]
+ \left( \Omega_n - \lambda_n \right)^2 \coth \left[\frac{\hbar \beta}{2} \left( \Omega_n - \lambda_n \right) \right] \right\}
\end{array}
\end{equation}
In the commutative limit we recover the expected results \cite{Halliwell}:
\begin{equation}
\left\{
\begin{array}{l}
< \hat p_i ^{(n)} \hat p_j ^{(m)}>  \longrightarrow \frac{\hbar m_n \omega_n \delta_{n,m} \delta_{i,j}}{2 } \coth \left( \frac{\hbar \beta \omega_n}{2} \right)\\
\\
< \frac{\hat q_i ^{(n)} \hat p_j ^{(m)} + \hat p_j ^{(m)} \hat q_i ^{(n)}}{2}>  \longrightarrow 0 \\
\\
< \frac{\hat q_i ^{(n)} \hat q_j ^{(m)} + \hat q_j ^{(m)} \hat q_i ^{(n)}}{2}>  \longrightarrow \frac{\hbar \delta_{n,m} \delta_{i,j}}{2 m_n \omega_n} \coth \left( \frac{\hbar \beta \omega_n}{2} \right)
\end{array}
\right.
\end{equation}

\section*{Appendix 2}

Our task is now to determine the coefficients ${\bf A}$, ${\bf B}$, ${\bf C}$, ${\bf D}$. From eq.(33) we have:
\begin{equation}
\begin{array}{l l}
\frac{d}{dt} < \hat q_i >  & = \frac{<\hat p_i>}{M} + \frac{\theta}{\hbar} M \Omega^2 \epsilon_{ij} <\hat q_j> + \frac{\theta}{\hbar}  \epsilon_{ij} \sum_n C_n <\hat q_j^{(n)}> \\
& \\
\frac{d}{dt} < \frac{\hat q_i \hat q_j + \hat q_j \hat q_i}{2}> & = \frac{1}{M} < \frac{\hat q_i \hat p_j + \hat p_j \hat q_i}{2}> + \frac{\theta}{\hbar} M \Omega^2 \epsilon_{ik} < \frac{\hat q_j \hat q_k + \hat q_k \hat q_j}{2}> + \frac{\theta}{\hbar}  \epsilon_{ik} \sum_n C_n < \hat q_j \hat q_k^{(n)}>  + \left( i \longleftrightarrow j \right)\\
& \\
\frac{d}{dt} < \frac{\hat p_i \hat q_j + \hat q_j \hat p_i}{2}> & = \frac{1}{M} < \hat p_i \hat p_j> -  M \Omega^2  < \frac{\hat q_i \hat q_j + \hat q_j \hat q_i}{2}> - \sum_n C_n < \hat q_i^{(n)}  \hat q_j> + \frac{\theta}{\hbar}  M \Omega^2 \epsilon_{jk} < \frac{\hat p_i \hat q_k + \hat q_k \hat p_i}{2}> + \\
& \\
& \hspace{0.4 cm} + \frac{\theta}{\hbar} \epsilon_{jk}  \sum_n C_n <   \hat p_i \hat q_k^{(n)}>
\end{array}
\end{equation}
Alternatively, we may perform the calculation using eq.(38) instead:
\begin{equation}
\begin{array}{l l}
\frac{d}{dt} < \hat q_i >  & = \frac{<\hat p_i>}{M} + \frac{\theta}{\hbar} M \Omega^2 \epsilon_{ij} <\hat q_j> + \frac{\theta}{\hbar}  \epsilon_{ij} A_{jk} <\hat q_k> + \frac{\theta}{\hbar}  \epsilon_{ij} B_{jk} <\hat p_k> \\
& \\
\frac{d}{dt} < \frac{\hat q_i \hat q_j + \hat q_j \hat q_i}{2}> & = \frac{1}{M} < \frac{\hat q_i \hat p_j + \hat p_j \hat q_i}{2}> + \frac{\theta}{\hbar} M \Omega^2 \epsilon_{ik} < \frac{\hat q_j \hat q_k + \hat q_k \hat q_j}{2}> + \frac{\theta}{\hbar}  \epsilon_{ik} A_{kl}  < \frac{\hat q_j \hat q_l + \hat q_l \hat q_j}{2}> + \\
& \\
& \hspace{0.4 cm} + \frac{\theta}{\hbar} \epsilon_{ik} B_{kl}  < \frac{\hat q_j \hat p_l + \hat p_l \hat q_j}{2}> + \frac{\theta}{\hbar}  \epsilon_{kj} C_{ki} + \left( i \longleftrightarrow j \right)\\
& \\
\frac{d}{dt} < \frac{\hat p_i \hat q_j + \hat q_j \hat p_i}{2}> & = \frac{1}{M} < \hat p_i \hat p_j> -  M \Omega^2  < \frac{\hat q_i \hat q_j + \hat q_j \hat q_i}{2}> - A_{ik} < \frac{\hat q_j \hat q_k + \hat q_k \hat q_j}{2}> - B_{ik} < \frac{\hat q_j \hat p_k + \hat p_k \hat q_j}{2}>+ C_{ij}\\
& \\
& \hspace{0.4 cm} + \frac{\theta}{\hbar} M \Omega^2 \epsilon_{jk} < \frac{\hat p_i \hat q_k + \hat q_k \hat p_i}{2}> + \frac{\theta}{\hbar} \epsilon_{jk} A_{kl} < \frac{\hat p_i \hat q_l + \hat q_l \hat p_i}{2}> + \frac{\theta}{\hbar} \epsilon_{jk} B_{kl} < \hat p_i \hat p_l > - \frac{\theta}{\hbar} \epsilon_{jk} D_{ki}
\end{array}
\end{equation}
Equating (126) and (127), we obtain the following relations:
\begin{equation}
\left\{
\begin{array}{l}
\sum_n C_n <\hat q_i^{(n)}> = A_{ij} <\hat q_j>  + B_{ij} <\hat p_j>\\
\\
M_{jk} = \frac{\theta}{\hbar} \epsilon_{jl} N_{lk} = M_{11} \delta_{jk}
\end{array}
\right.
\end{equation}
where:
\begin{equation}
\left\{
\begin{array}{l}
M_{jk} \equiv \sum_n C_n <\hat q_j^{(n)} \hat q_k > -A_{jl}  <\frac{\hat q_k \hat q_l + \hat q_l \hat q_k }{2}> - B_{jl}  <\frac{\hat q_k \hat p_l + \hat p_l \hat q_k }{2}>  + C_{jk} =  \Lambda_{jk}+ C_{jk}\\
\\
N_{jk} \equiv \sum_n C_n <\hat q_j^{(n)} \hat p_k > -A_{jl}  <\frac{\hat p_k \hat q_l + \hat q_l \hat p_k }{2}> - B_{jl}  <\hat p_k \hat p_l>  + D_{jk}  = \Omega_{jk}+ D_{jk}
\end{array}
\right.
\end{equation}
Now let us go back to eq.(38) and consider all the terms that include contributions from the matrices ${\bf C}$ and ${\bf D}$:
\begin{equation}
\begin{array}{c}
\left(C_{ij} - \frac{\theta}{\hbar} \epsilon_{jk} D_{ki} \right) \frac{\partial^2 W_r}{\partial p_i \partial q_j} + D_{ij} \frac{\partial^2 W_r}{\partial p_i \partial p_j} + \frac{\theta}{\hbar} \epsilon_{ij} C_{ik} \frac{\partial^2 W_r}{\partial q_j \partial q_k} =\\
\\
= \left[ \left(M_{ij} - \frac{\theta}{\hbar} \epsilon_{jk} N_{ki} \right) - \Lambda_{ij} + \frac{\theta}{\hbar} \epsilon_{jk} \Omega_{ki} \right] \frac{\partial^2 W_r}{\partial p_i \partial q_j}
+ \left( N_{ij} - \Omega_{ij} \right) \frac{\partial^2 W_r}{\partial p_i \partial p_j} + \frac{\theta}{\hbar} \epsilon_{ij}\left( M_{ik} - \Lambda_{ik} \right) \frac{\partial^2 W_r}{\partial q_j \partial q_k} = \\
\\
=  \left( - \Lambda_{ij} + \frac{\theta}{\hbar} \epsilon_{jk} \Omega_{ki} \right) \frac{\partial^2 W_r}{\partial p_i \partial q_j} -  \Omega_{ij}  \frac{\partial^2 W_r}{\partial p_i \partial p_j} - \frac{\theta}{\hbar} \epsilon_{ij} \Lambda_{ik}\frac{\partial^2 W_r}{\partial q_j \partial q_k},
\end{array}
\end{equation}
where we used (129) and the fact that ${\bf M}$ and ${\bf N}$ are symmetric and antisymmetric, respectively. Since the matrices ${\bf M}$, ${\bf N}$ do not contribute to our equation it is perfectly consistent to set them to zero. Altogether we have:
\begin{equation}
\left\{
\begin{array}{l}
\sum_n C_n <\hat q_i^{(n)}> = A_{ij}  < \hat q_j> + B_{ij}  <\hat p_j> \\
\\
\sum_n C_n <\hat q_j^{(n)} \hat q_k > = A_{jl}  <\frac{\hat q_k \hat q_l + \hat q_l \hat q_k }{2}> + B_{jl}  <\frac{\hat q_k \hat p_l + \hat p_l \hat q_k }{2}>  - C_{jk} \\
\\
\sum_n C_n <\hat q_j^{(n)} \hat p_k > = A_{jl}  <\frac{\hat p_k \hat q_l + \hat q_l \hat p_k }{2}> + B_{jl}  <\hat p_k \hat p_l>  - D_{jk}
\end{array}
\right.
\end{equation}
The coefficients ${\bf A}$, ${\bf B}$, ${\bf C}$, ${\bf D}$ may now be determined from the previous equation once we have computed the expectation values $<\hat q_i>$, $<\hat q_i^{(n)} \hat q_j>$, etc of the Heisenberg picture operators by solving the equations of motion. Since the Hamiltonian is quadratic we have the following linear solutions:
\begin{equation}
\left\{
\begin{array}{l}
\hat q_i^{(n)} (t) = \alpha_{ij}^{(n)} (t) \hat q_j (t)  + \beta_{ij}^{(n)} (t) \hat p_j (t) + \sum_m \left( a_{ij}^{(n,m)} (t) \hat q_j^{(m)} (0) + b_{ij}^{(n,m)} (t) \hat p_j^{(m)} (0) \right) \\
\\
\hat q_i (t) = \alpha_{ij} (t) \hat q_j (0)  + \beta_{ij} (t) \hat p_j (0) + \sum_m \left( a_{ij}^{(m)} (t) \hat q_j^{(m)} (0) + b_{ij}^{(m)} (t) \hat p_j^{(m)} (0) \right)
\end{array}
\right.
\end{equation}
for some time dependent coefficients $\alpha_{ij}^{(n)}$, $\beta_{ij}^{(n)}$, etc. Notice that we have chosen to express $\hat q_i^{(n)}$ in terms of $\hat q_j (t)$ and $\hat p_j (t)$, instead of $\hat q_j (0)$ and $\hat p_j (0)$. This is for later convenience \cite{Halliwell}. If  we substitute (132) in (131) by keeping (30) in mind, we obtain:
\begin{equation}
\begin{array}{l}A_{ij} (t) = \sum_n C_n \alpha_{ij}^{(n)} (t), \hspace{0.5 cm} B_{ij} (t) = \sum_n C_n \beta_{ij}^{(n)} (t) \\
\\
C_{ij} (t) = - \sum_{n,m} C_n \left\{a_{il}^{(n,m)} \left[a_{jl}^{(m)} <\left[\hat q_l^{(m)} (0) \right]^2 > +  b_{jk}^{(m)} <\frac{\hat p_k^{(m)} (0) \hat q_l^{(m)} (0) + \hat q_l^{(m)} (0) \hat p_k^{(m)} (0)}{2} > \right] + \right.\\
\\
\left. \hspace{1.3 cm} +  b_{il}^{(n,m)} \left[a_{jk}^{(m)} <\frac{\hat q_k^{(m)} (0) \hat p_l^{(m)} (0) + \hat p_l^{(m)} (0) \hat q_k^{(m)} (0)}{2} > +  b_{jl}^{(m)} <\left[\hat p_l^{(m)} (0) \right]^2 > \right] \right\}\\
\\
D_{ij} (t) = - M \sum_{n,m} C_n \left\{a_{ik}^{(n,m)} \left[\dot a_{jk}^{(m)} <\left[\hat q_k^{(m)} (0) \right]^2 > +  \dot b_{jl}^{(m)} <\frac{\hat q_k^{(m)} (0) \hat p_l^{(m)} (0) + \hat p_l^{(m)} (0) \hat q_k^{(m)} (0)}{2} > \right] + \right.\\
\\
\left. \hspace{1.3 cm} +  b_{ik}^{(n,m)} \left[\dot a_{jl}^{(m)} <\frac{\hat p_k^{(m)} (0) \hat q_l^{(m)} (0) + \hat q_l^{(m)} (0) \hat p_k^{(m)} (0)}{2} > +  \dot b_{jk}^{(m)} <\left[\hat p_k^{(m)} (0) \right]^2 > \right] \right\}
\end{array}
\end{equation}
From (132,133) the problem is solved once we have the solutions of the equations of motion for the Heisenberg picture operators. For the sake of simplicity we shall omit the hats over the operators. There is no risk of confusion here as all variables appearing in the ensuing analysis are operators. The remaining calculations are lengthy and involved. To keep the task as tractable as possible we have chosen to follow ref.\cite{Halliwell} step-by-step and to adopt their notation.

To solve the equations of motion it turns out to be more convenient to consider the following set of variables:
\begin{equation}
\left\{
\begin{array}{l}
Q_i = q_i + \frac{\theta}{\hbar} \epsilon_{ij} p_j\\
P_i = p_i
\end{array}
\right.
\hspace{1 cm}
\left\{
\begin{array}{l}
Q_i^{(n)} = q_i^{(n)} + \frac{\theta}{\hbar} \epsilon_{ij} p_j^{(n)}\\
P_i^{(n)} = p_i^{(n)}
\end{array}
\right.
\end{equation}
The {\it rationale} for this choice resides in the fact that
\begin{equation}
P_i = M \dot Q_i, \hspace{1 cm} P_i^{(n)} = m_n \dot Q_i^{(n)}.
\end{equation}
We then get the following set of equations of motion:
\begin{equation}
\left\{
\begin{array}{l}
\ddot Q_i + \Omega^2 Q_i  + \frac{1}{M} \sum_n C_n Q_i^{(n)} - \frac{\theta}{\hbar} M \Omega^2 \epsilon_{ij} \dot Q_j - \frac{\theta}{\hbar} \sum C_n \frac{m_n}{M} \epsilon_{ij} \dot Q_j^{(n)}=0\\
\\
\ddot Q_i^{(n)} + \omega_n^2 Q_i^{(n)}  + \frac{C_n}{m_n} Q_i - \frac{\theta}{\hbar} m_n \omega_n^2 \epsilon_{ij} \dot Q_j^{(n)} - \frac{\theta}{\hbar} C_n \frac{M}{m_n} \epsilon_{ij} \dot Q_j=0
\end{array}
\right.
\end{equation}
We start by solving the latter equation. The homogeneous equation reads:
\begin{equation}
\ddot Q_i^{(n)} + \omega_n^2 Q_i^{(n)} - \frac{\theta}{\hbar} m_n \omega_n^2 \epsilon_{ij} \dot Q_j^{(n)} =0
\end{equation}
Upon diagonalization we obtain the following solution:
\begin{equation}
Q_i^{(n)} (t) = \lambda_{ij}^{(n)} (t) Q_j^{(n)} (0) + \rho_{ij}^{(n)} (t) \frac{P_j^{(n)} (0)}{m_n},
\end{equation}
subject to the initial conditions:
\begin{equation}
Q_i^{(n)} (t=0) = Q_i^{(n)} (0), \hspace{1 cm} \dot Q_i^{(n)} (t=0) = \frac{P_i^{(n)} (0)}{m_n}.
\end{equation}
The matrices $\lambda^{(n)}$, $\rho^{(n)}$ are given by:
\begin{equation}
\left\{
\begin{array}{l}
\lambda_{ij}^{(n)} (t) = \frac{1}{2 \Omega_n} \sum_{\sigma= \pm} \left\{ \delta_{ij} \left( \Omega_n - \sigma \lambda_n  \right) \cos \left[ \left( \Omega_n + \sigma \lambda_n \right)t \right] + \sigma \epsilon_{ij} \left( \Omega_n - \sigma \lambda_n  \right) \sin \left[ \left( \Omega_n + \sigma \lambda_n \right)t \right] \right\}\\
\\
\rho_{ij}^{(n)} (t) = \frac{1}{2 \Omega_n} \sum_{\sigma= \pm} \left\{ \delta_{ij}  \sin \left[ \left( \Omega_n + \sigma \lambda_n \right)t \right] - \sigma \epsilon_{ij}  \cos \left[ \left( \Omega_n + \sigma \lambda_n \right)t \right] \right\}
\end{array}
\right.
\end{equation}
where $\Omega_n$, $\lambda_n$ are given by (27). It is useful to remark that:
\begin{equation}
\dot \lambda_{ij}^{(n)} = - \omega_n^2 \rho_{ij}^{(n)}, \hspace{1 cm} \dot \rho_{ij}^{(n)}= \lambda_{ij}^{(n)} + 2 \lambda_n \epsilon_{ik} \rho_{kj}^{(n)}.
\end{equation}
The matrix $\lambda^{(n)}$ is a solution of the differential equation:
\begin{equation}
\ddot \lambda_{ij}^{(n)} + \omega_n^2 \lambda_{ij}^{(n)} - 2 \lambda_n \epsilon_{ik} \dot \lambda_{kj}^{(n)} =0.
\end{equation}
Because of relation (141), $\rho^{(n)}$ is a solution of the same equation. Finally, they obey the following initial conditions (cf.(139)):
\begin{equation}
\left\{
\begin{array}{l l}
\lambda_{ij}^{(n)} (t=0) = \delta_{ij}  & \rho_{ij}^{(n)} (t=0) = 0\\
\\
\dot \lambda_{ij}^{(n)} (t=0) = 0 & \dot \rho_{ij}^{(n)} (t=0) = \delta_{ij}
\end{array}
\right.
\end{equation}
A particular solution of the second inhomogeneous equation in (136) is given by:
\begin{equation}
g_i^{(n)} (t) = - \frac{C_n}{m_n} \int_0^t ds \hspace{0.2 cm} \rho_{kj}^{(n)} (t-s) \left(\delta_{ik} -  \frac{M \theta}{\hbar} \epsilon_{ik} \frac{d}{ds} \right) Q_j (s),
\end{equation}
which satisfies the initial conditions:
\begin{equation}
g_i^{(n)} (t=0) = \dot g_i^{(n)} (t=0) =0.
\end{equation}
Altogether, the complete solution of the second equation in (136) subject to the conditions (139) is:
\begin{equation}
Q_i^{(n)} (t) = \lambda_{ij}^{(n)}  (t) Q_j^{(n)} (0) + \rho_{ij}^{(n)}  (t) \frac{P_j^{(n)} (0)}{m_n} + g_i^{(n)} (t).
\end{equation}
If we substitute this expression in the first of eqs.(136) we obtain after some algebra the noncommutative Langevin equation:
\begin{equation}
\ddot Q_i (t) + \Omega^2 Q_i (t) - \frac{\theta}{\hbar} M \Omega^2 \epsilon_{ij} \dot Q_j (t) + \frac{2}{M} \int_0^t ds \hspace{0.2 cm} \eta_{kj} (t-s) \left(\delta_{ik} - \frac{M \theta}{\hbar} \epsilon_{ik} \frac{d}{ds} \right) Q_j (s) = \frac{f_i (t)}{M},
\end{equation}
The inhomogeneous term in (147) is:
\begin{equation}
f_i (t) = - \sum_n C_n \left\{ \left[ \lambda_{ij}^{(n)} (t) + 2 \lambda_n \epsilon_{ik} \rho_{kj}^{(n)}(t) \right] Q_j^{(n)} (0) + \left[ \left(1+ \frac{4 \lambda_n^2}{\omega_n^2} \right) \rho_{ij}^{(n)} (t) - \frac{2 \lambda_n}{\omega_n^2} \epsilon_{ik} \lambda_{kj}^{(n)}(t) \right] \frac{P_j^{(n)} (0)}{m_n} \right\}.
\end{equation}
Equation (147) describes the motion of a noncommutative Brownian particle in a dissipative medium with "random" force $f_i (t)$ satisfying (cf.(30,40)):
\begin{equation}
<f_i (t)> =0, \hspace{1 cm} < \left\{f_i (t), f_j (t') \right\}>= \hbar \nu_{ij} (t -t'),
\end{equation}
where $\left\{A,B \right\} = \left( AB +BA \right)/2$ is the anticommutator. Notice that in the commutative limit $\theta \to 0$, eqs.(41,140,148) reduce to:
\begin{equation}
\left\{
\begin{array}{l}
\lambda_{ij}^{(n)} (t) \longrightarrow \delta_{ij} \cos (\omega_n t), \hspace{1 cm} \rho_{ij}^{(n)} (t) \longrightarrow \frac{\delta_{ij}}{\omega_n} \sin (\omega_n t)\\
\\
I^+ (\omega) \longrightarrow  \sum_n \frac{C_n^2}{2 m_n \omega_n} \delta (\omega -\omega_n) , \hspace{1 cm} I^- (\omega) \longrightarrow 0\\
\\
f_i (t) \longrightarrow - \sum_n C_n \left[Q_i^{(n)} (0) \cos (\omega_n t) + \frac{P_i^{(n)} (0)}{m_n} \frac{\sin (\omega_n t)}{\omega_n} \right],
\end{array}
\right.
\end{equation}
which are the expected results \cite{Halliwell}. Our aim is to solve eq.(147). As in \cite{Halliwell}, we solve it for two different sets of initial conditions\footnote{As in \cite{Halliwell} $s$ is henceforth the variable and $t$ a specific fixed time. This is related with the fact that in eq. (132) we expressed $q_i^{(n)} (t)$ in terms of $q_i (t),p_i (t)$ and not $q_i (0),p_i (0)$.}:
\begin{equation}
Q_i (s=0) = Q_i (0), \hspace{1 cm} \dot Q_i (s=0) = \frac{P_i (0)}{M},
\end{equation}
and
\begin{equation}
Q_i (s=t) = Q_i (t), \hspace{1 cm} \dot Q_i (s=t) = \frac{P_i (t)}{M},
\end{equation}
where $t >0$ is any given time. The solution of the homogeneous equation $(f_i (t) =0)$ (eq.(147)) is given by:
\begin{equation}
\omega_i (s) = \left[ u_{ij} (s) - v_{ik} (s) \left(\dot v (0)\right)^{-1}_{kl}  \dot u_{lj} (0) \right] Q_j (0) + v_{ik} (s) \left(\dot v (0)\right)^{-1}_{kj}\frac{P_j (0)}{M}.
\end{equation}
The particular solution to eq.(147) with initial conditions $\tilde{\omega}_i (s=0) = \dot{\tilde{\omega}}_i (s=0)=0$, can be formally written as:
\begin{equation}
\tilde{\omega}_i (s) = \frac{1}{M}\int_0^s d \tau \hspace{0.2 cm} G_{ij}^{(1)} (s, \tau ) f_j ( \tau),
\end{equation}
where $G^{(1)} (s , \tau)$ is the Green function (44). The solution of (147) with the conditions (151) is then:
\begin{equation}
\begin{array}{c}
Q_i (s) = \omega_i (s) + \tilde{\omega}_i (s) =\left[ u_{ij} (s) - v_{ik} (s) \left(\dot v (0)\right)^{-1}_{kl}  \dot u_{lj} (0) \right] Q_j (0) + v_{ik} (s) \left(\dot v (0)\right)^{-1}_{kj}\frac{P_j (0)}{M} \\
\\
 - \sum_n \frac{C_n}{M}  \int_0^s d \tau \hspace{0.2 cm} G_{ij}^{(1)} (s, \tau) \left[ \lambda_{jk}^{(n)} ( \tau) + 2 \lambda_n \epsilon_{jl} \rho_{lk}^{(n)} (\tau) \right] Q_k^{(n)} (0)\\
\\
 - \sum_n \frac{C_n}{M}  \int_0^s d \tau \hspace{0.2 cm} G_{ij}^{(1)} (s, \tau) \left[ \left(1 + \frac{4 \lambda_n^2}{\omega_n^2} \right) \rho_{jk}^{(n)} ( \tau) - \frac{2 \lambda_n}{\omega_n^2} \epsilon_{jl} \lambda_{lk}^{(n)} (\tau) \right] \frac{P_k^{(n)} (0)}{m_n}.
\end{array}
\end{equation}
Likewise, from (46,135,155) we get:
\begin{equation}
\begin{array}{c}
P_i (s) = M \left[ \dot u_{ij} (s) - \dot v_{ik} (s) \left(\dot v (0)\right)^{-1}_{kl}  \dot u_{lj} (0) \right] Q_j (0) + \dot v_{ik} (s) \left(\dot v (0)\right)^{-1}_{kj}P_j (0) \\
\\
 - \sum_n C_n  \int_0^s d \tau \hspace{0.2 cm} \left[\frac{d}{ds} G_{ij}^{(1)} (s, \tau) \right]  \left[ \lambda_{jk}^{(n)} ( \tau) + 2 \lambda_n  \epsilon_{jl} \rho_{lk}^{(n)} (\tau) \right] Q_k^{(n)} (0)\\
\\
 - \sum_n C_n \int_0^s d \tau \hspace{0.2 cm} \left[\frac{d}{ds} G_{ij}^{(1)} (s, \tau) \right]   \left[ \left(1 + \frac{4 \lambda_n^2}{\omega_n^2} \right) \rho_{jk}^{(n)} ( \tau) - \frac{2 \lambda_n}{\omega_n^2}  \epsilon_{jl} \lambda_{lk}^{(n)} (\tau) \right] \frac{P_k^{(n)} (0)}{m_n},
\end{array}
\end{equation}
Next, we look for the solutions of the homogeneous eq.(147) $(f_i (t)=0)$ with initial conditions (152). The result is:
\begin{equation}
Z_i (s) = \left[ v_{ij} (s) - u_{ik} (s) \left(\dot u (t)\right)^{-1}_{kl}  \dot v_{lj} (t) \right] Q_j (t) + u_{ik} (s) \left(\dot u (t)\right)^{-1}_{kj} \frac{P_j (t)}{M}.
\end{equation}
The solution of the inhomogeneous eq.(147) with the conditions $\tilde Z_i (s=t) = \dot{\tilde Z}_i (s=t)=0$, is given by:
\begin{equation}
\tilde Z_i (s) = \frac{1}{M} \int_t^s d \tau \hspace{0.2 cm} G_{ij}^{(2)} (s, \tau ) f_j ( \tau), \hspace{0.5 cm} (s \le t )
\end{equation}
where $G^{(2)} (s)$ is the Green function (44). Altogether, the solution of (147) with the conditions (152) is then:
\begin{equation}
\begin{array}{c}
Q_i (s) = Z_i (s) + \tilde Z_i (s) =  \left[ v_{ij} (s) - u_{ik} (s) \left(\dot u (t)\right)^{-1}_{kl}  \dot v_{lj} (t) \right] Q_j (t) + u_{ik} (s) \left(\dot u (t)\right)^{-1}_{kj} \frac{P_j (t)}{M} + \\
\\
 + \sum_n \frac{C_n}{M} \int_s^t d \tau \hspace{0.2 cm} G_{ij}^{(2)} (s, \tau) \left[ \lambda_{jk}^{(n)} ( \tau) + 2 \lambda_n  \epsilon_{jl} \rho_{lk}^{(n)} (\tau) \right] Q_k^{(n)} (0) +\\
\\
 + \sum_n \frac{C_n}{M}  \int_s^t d \tau \hspace{0.2 cm} G_{ij}^{(2)} (s, \tau) \left[ \left(1 + \frac{4 \lambda_n^2}{\omega_n^2} \right) \rho_{jk}^{(n)} ( \tau) - \frac{2 \lambda_n}{\omega_n^2} \epsilon_{jl} \lambda_{lk}^{(n)} (\tau) \right] \frac{P_k^{(n)} (0)}{m_n}.
\end{array}
\end{equation}
From (46,135,159) we get:
\begin{equation}
\begin{array}{c}
P_i (s) = M  \left[ \dot v_{ij} (s) - \dot u_{ik} (s) \left(\dot u (t)\right)^{-1}_{kl}  \dot v_{lj} (t) \right] Q_j (t) + \dot u_{ik} (s) \left(\dot u (t)\right)^{-1}_{kj} P_j (t) + \\
\\
 + \sum_n C_n \int_s^t d \tau \hspace{0.2 cm} \left[\frac{d}{ds} G_{ij}^{(2)} (s, \tau) \right]  \left[ \lambda_{jk}^{(n)} ( \tau) + 2 \lambda_n  \epsilon_{jl} \rho_{lk}^{(n)} (\tau) \right] Q_k^{(n)} (0) +\\
\\
 + \sum_n C_n  \int_s^t d \tau \hspace{0.2 cm} \left[\frac{d}{ds} G_{ij}^{(2)} (s, \tau) \right]   \left[ \left(1 + \frac{4 \lambda_n^2}{\omega_n^2} \right) \rho_{jk}^{(n)} ( \tau) - \frac{2 \lambda_n}{\omega_n^2}  \epsilon_{jl} \lambda_{lk}^{(n)} (\tau) \right] \frac{P_k^{(n)} (0)}{m_n}.
\end{array}
\end{equation}
Substitution of (159) in (144,146) yields:
\begin{equation}
\begin{array}{c}
Q_i^{(n)} (t) =  \lambda_{ij}^{(n)} (t) Q_j^{(n)} (0) +  \rho_{ij}^{(n)} (t) \frac{P_j^{(n)} (0)}{m_n}\\
\\
- \frac{C_n}{m_n}  \int_0^t ds \hspace{0.2 cm} \rho_{kj}^{(n)} (t-s) \left( \delta_{ik} - \frac{M \theta}{\hbar} \epsilon_{ik} \frac{d}{ds} \right) \left\{  \left[ v_{jl} (s) - u_{ja} (s) \left(\dot u (t)\right)^{-1}_{ab}  \dot v_{bl} (t) \right] Q_l (t) + u_{ja} (s) \left(\dot u (t)\right)^{-1}_{al} \frac{P_l (t)}{M}+  \right. \\
\\
+ \sum_m \frac{C_m}{M}  \int_s^t d \tau \hspace{0.2 cm}  G_{ja}^{(2)} (s, \tau)  \left[ \lambda_{al}^{(m)} ( \tau) + 2 \lambda_m  \epsilon_{ar} \rho_{rl}^{(m)} (\tau) \right] Q_l^{(m)} (0) +\\
\\
\left. + \sum_m \frac{C_m}{M}  \int_s^t d \tau \hspace{0.2 cm}  G_{ja}^{(2)} (s, \tau)    \left[ \left(1 + \frac{4 \lambda_m^2}{\omega_m^2} \right) \rho_{al}^{(m)} ( \tau) - \frac{2 \lambda_m}{\omega_m^2} \epsilon_{ar} \lambda_{rl}^{(m)} (\tau) \right] \frac{P_l^{(m)}  (0)}{m_m} \right\} .
\end{array}
\end{equation}
And similarly:
\begin{equation}
\begin{array}{c}
P_i^{(n)} (t) =  m_n \dot \lambda_{ij}^{(n)} (t) Q_j^{(n)} (0) +  \dot \rho_{ij}^{(n)} (t) P_j^{(n)} (0)\\
\\
- C_n  \int_0^t ds \hspace{0.2 cm} \dot \rho_{kj}^{(n)} (t-s) \left( \delta_{ik} - \frac{M \theta}{\hbar} \epsilon_{ik} \frac{d}{ds} \right) \left\{  \left[ v_{jl} (s) - u_{ja} (s) \left(\dot u (t)\right)^{-1}_{ab}  \dot v_{bl} (t) \right] Q_l (t) + u_{ja} (s) \left(\dot u (t)\right)^{-1}_{al} \frac{P_l (t)}{M}+  \right. \\
\\
+ \sum_m \frac{C_m}{M}  \int_s^t d \tau \hspace{0.2 cm}  G_{ja}^{(2)} (s, \tau)  \left[ \lambda_{al}^{(m)} ( \tau) + 2 \lambda_m  \epsilon_{ar} \rho_{rl}^{(m)} (\tau) \right] Q_l^{(m)} (0) +\\
\\
\left. + \sum_m \frac{C_m}{M}  \int_s^t d \tau \hspace{0.2 cm}  G_{ja}^{(2)} (s, \tau)    \left[ \left(1 + \frac{4 \lambda_m^2}{\omega_m^2} \right) \rho_{al}^{(m)} ( \tau) - \frac{2 \lambda_m}{\omega_m^2} \epsilon_{ar} \lambda_{rl}^{(m)} (\tau) \right] \frac{P_l^{(m)}  (0)}{m_m} \right\} .
\end{array}
\end{equation}
Using these expressions as well as the relations (133,134), we obtain after some manipulations:
\begin{equation}
\begin{array}{l l}
\alpha_{ij}^{(n)} (t) = & - \frac{C_n}{m_n}  \int_0^t ds \hspace{0.2 cm} \left[ \left(1 + \frac{4 \lambda_n^2}{\omega_n^2} \right) \rho_{kl}^{(n)} (t-s) - \frac{2 \lambda_n}{\omega_n^2} \epsilon_{kr} \lambda_{rl}^{(n)} (t-s) \right] \left( \delta_{ik} - \frac{M \theta}{\hbar} \epsilon_{ik} \frac{d}{ds} \right) \left[ v_{lj} (s) - u_{la} (s) \left(\dot u (t)\right)^{-1}_{ab}  \dot v_{bj} (t) \right]\\
& \\
\beta_{ij}^{(n)} (t) = & - \frac{C_n}{m_n M}  \int_0^t ds \hspace{0.2 cm}
 \left[ \left(1 + \frac{4 \lambda_n^2}{\omega_n^2} \right) \rho_{lk}^{(n)} (t-s) - \frac{2 \lambda_n}{\omega_n^2} \epsilon_{lr} \lambda_{rk}^{(n)} (t-s) \right] \left( \delta_{il} - \frac{M \theta}{\hbar} \epsilon_{il} \frac{d}{ds} \right) \times \\
& \\
& \times \left\{u_{ka} (s) \left(\dot u (t)\right)^{-1}_{aj} + \frac{\theta M}{\hbar} \epsilon_{rj} \left[v_{kr} (s) - u_{ka} (s) \left(\dot u (t)\right)^{-1}_{ab}  \dot v_{br} (t) \right] \right\}
\end{array}
\end{equation}
And also:
\begin{equation}
\begin{array}{l l}
a_{ij}^{(n,m)} (t) = & \left[ \lambda_{ij}^{(n)} (t) +  2 \lambda_n \epsilon_{ik} \rho_{kj}^{(n)} (t) \right] \delta_{n,m} - \frac{C_n C_m}{m_n M} \int_0^t ds \int_s^t d \tau \hspace{0.2 cm} \left[ \lambda_{kj}^{(m)} (\tau) +  2 \lambda_m  \epsilon_{kr} \rho_{rj}^{(m)} (\tau) \right] \times\\
& \\
& \times \Lambda_{lk}^{(2)} (s, \tau) \left[ \left(1 + \frac{4 \lambda_n^2}{\omega_n^2} \right) \rho_{il}^{(n)} (t-s) - \frac{2 \lambda_n}{\omega_n^2}\epsilon_{ip} \lambda_{pl}^{(n)} (t-s) \right]\\
& \\
b_{ij}^{(n,m)} (t) = & \frac{\rho_{ij}^{(n)} (t)}{m_n}\delta{n,m} - \frac{C_n C_m}{m_n m_m M} \int_0^t ds \int_s^t d \tau \hspace{0.2 cm} \rho_{kj}^{(m)} (\tau) \Lambda_{lk}^{(2)} (s, \tau) \times \\
& \\
&  \times \left[ \left(1 + \frac{4 \lambda_n^2}{\omega_n^2} \right) \rho_{il}^{(n)} (t-s) - \frac{2 \lambda_n}{\omega_n^2}  \epsilon_{ip} \lambda_{pl}^{(n)} (t-s) \right]
\end{array}
\end{equation}
Moreover:
\begin{equation}
\begin{array}{l l}
a_{ij}^{(n)} (t) = & - \frac{C_n}{M} \int_0^t ds  \hspace{0.2 cm} \left[ \lambda_{kj}^{(n)} (s) +  2 \lambda_n  \epsilon_{kl} \rho_{lj}^{(n)} (s) \right]\Lambda_{ik}^{(1)} (t,s) \\
& \\
b_{ij}^{(n)} (t) = & - \frac{C_n}{m_n M} \int_0^t ds  \hspace{0.2 cm}  \rho_{kj}^{(n)} (s) \Lambda_{ik}^{(1)} (t,s)
\end{array}
\end{equation}
Finally, if we substitute eqs.(163-165) in (133), we recover (47,48).

\begin{center}

{\large{{\bf Acknowledgments}}}

\end{center}

\vspace{0.3 cm}
\noindent
The authors wish to thank O. Bertolami and A. Mikovic for useful comments and for reading the manuscript. This work was partially supported by the grants POCTI/MAT/45306/2002 and POCTI/0208/2003 of the Portuguese Science Foundation.

\end{document}